\documentclass{article}

\usepackage[english]{babel}
\usepackage{natbib}

\usepackage[letterpaper,top=2cm,bottom=2cm,left=3cm,right=3cm,marginparwidth=1.75cm]{geometry}
\usepackage[colorlinks=true,linkcolor=blue,citecolor=blue]{hyperref}
\usepackage{framed}
\usepackage{float}
\usepackage{bm}
\usepackage{multirow}
\usepackage{amsmath}
\usepackage{graphicx}
\usepackage{amsfonts}
\usepackage{amssymb}
\usepackage{enumerate}
\usepackage{float}
\usepackage{xcolor}
\usepackage{booktabs}
\usepackage{bbm}
\usepackage{listings}
\usepackage{pdflscape}

\title{
Bayesian Multivariate Approach to Subnational mortality graduation with Age-Varying Smoothness}
\author{Luiz F. V. de Figueiredo$^{2}$, \hspace{0.1cm} Viviana G. R. Lobo$^{1,2}$\footnote{{{\it Corresponding author}: Viviana G R Lobo, Departamento de M\'etodos Estat\'{\i}sticos, Instituto de Matem\'atica, Universidade Fe\-de\-ral do Rio de Janeiro, Av. Athos da Silveira Ramos, Centro de Tecnologia, Bloco C, CEP 21941-909. \newline {\it E-mail}: {\tt viviana@dme.ufrj.br}. {\it Homepage}: \url{https://sites.google.com/a/dme.ufrj.br/viviana/}}}, \\ Mariane B. Alves$^{1,2}$  \hspace{0.1cm} and  \hspace{0.1cm} Thais C. O. Fonseca$^{1,2}$ \\
\textit{$^{1}$Instituto de Matem\'atica, Departamento de Métodos Estatísticos} \\ 
\textit{$^{2}$Instituto de Matem\'atica, Laborat\'orio de Matem\'atica Aplicada} \\ 
\textit{Universidade Federal do Rio de Janeiro, Brazil }
}
\date{}

\begin{document}
\maketitle

\begin{abstract}
This work introduces a Bayesian smoothing approach for the joint graduation of mortality rates across multiple populations. In particular, dynamical linear models (DLMs) are used to induce smoothness across ages through structured dependence, analogously to how temporal correlation is accommodated in state-space time-indexed models.  An essential issue in subnational mortality probabilistic modelling is the lack or sparseness of information for some subpopulations considered in the analysis. For many countries, mortality data is severely limited, and approaches based on a single population model could result in a high level of uncertainty associated with the adjusted mortality tables. Here, we recognize the interdependence within a group of mortality data and pursue the approach of pooling information across several curves that ideally share common characteristics, such as the influence of epidemics and large economic changes. Our proposal considers multivariate Bayesian dynamical models with common parameters allowing for borrowing information across mortality tables. This will also allow for testing the hypothesis of convergence of several populations.  Furthermore, we employ discount factors, which are typical in DLMs, to regulate the smoothness of the fit, with varying discounting across the age domain. This ensures less smoothness at younger ages, while allowing for greater stability and slower changes in the mortality curve at adult ages. This setup implies a trade-off between stability and adaptability of mortality graduation. In particular, the discount parameter can be interpreted as a controller for the responsiveness of the fit for older age,  relative to new data. The estimation approach is fully Bayesian, accommodating all uncertainties in the modelling and prediction. To illustrate the effectiveness of our proposed model, we analyse the mortality dataset for males and females from England and Wales between 2010 and 2012, obtained from the Office for National Statistics database, across several scenarios.  Moreover, in the context of simulated missing data, our approach demonstrated good properties and flexibility in the pooling of information from other tables where data are available or patterns are clearer.

\end{abstract}
\vspace{0.3cm}

\section{Introduction}\label{sec1}

Mortality tables are often constructed in several areas such as government management, insurance product development and health studies. Modelling of empirical mortality rates usually considers a smooth function plus observational error. Life tables are then derived from the mortality laws inferred from the smooth model. Smoothing methods have been widely used in the literature and often consider parametric representation of mortality characteristics such as the Gompertz law \cite{gompertz1825xxiv}, the Makeham law \cite{makeham1860law} and the Heligman and Pollard's function \cite{heligman1980age}. Usually, laws of mortality consider sums or convolution of mathematical functions applicable to the infant, young adult, or/and adult ages. Ideally, interpretable parameters are included in these functions to capture flexible behaviours at specific age windows. However, estimation of model parameters could be problematic due to over-parameterization and/or high correlation between the parameters. 

Recently, other flexible smoothing approaches have been considered to model mortality, such as non-parametric smoothing approaches via splines. \cite{Eilers2004} propose a robust statistical framework for smoothing and forecasting mortality rates using penalized regression splines. The methods account for age and time variations, providing reliable and interpretable results that can be extended for future predictions.  \cite{Carmada2016} consider a sum of smooth exponentials and model mortality using shape-constrained P-splines and then consider smooth shape constraints for mortality forecast \cite{Carmada2019}. \cite{Forster2018} apply penalized splines and generalized additive models (GAMs) to smooth mortality data for the English Life Table 2010-2012, achieving improved accuracy and reliability in mortality estimates, while \cite{forster22} propose a model to smooth mortality rates based on a P-spline with adaptive penalty to allow for varying smoothness over the age domain.  An exhaustive reference review for splines in the mortality context can be seen in \cite{Eilers1996}, \cite{Camarda2008} and \cite{forster22}.

{If data is available over time, mortality rates might be modelled considering a stochastic model for the log-rates assuming the well-known Lee and Carter model \citep{Lee92}. This was a pioneer work in the mortality modelling of a single population. The method is based on a factor model with a latent factor varying across time and coefficients for each age allowing for estimation of improvement over the selected temporal window. Several multivariate extensions have been proposed to the Lee-Carter model. \cite{LiLee05} considers a joint model for closely related populations. The goal is to improve predictions of future mortality by including similar patterns for the group in the model. However, their proposal is based on the assumption that, in the long run, mortality levels for similar populations should not diverge. As a solution to guarantee convergence, they suggest that if the improvement parameters do not differ significantly from one population to the other, then it is reasonable to consider a common coefficient that ensures convergence \citep{lee1993modeling}. \cite{Dowd11} discuss a joint model for two related populations from a frequentist point of view. The authors consider the mortality modelling of a reference population with rich data and a population with sparseness through a gravitational model. In particular, Wales and England were considered related populations in the model. Other examples of joint models for two populations are \cite{cairns11} and references therein.}

Although there are many well-established proposals   to model a single population, the dependency between multiple populations is often misrepresented in the joint graduation process. As a result, populations with small exposure and mortality counts will have unreliable estimated tables with too wide uncertainty intervals. Mortality data for multi-populations can have some similarities in their dynamism that should be taken into account. For example, groups such as sex, educational level and regions usually present similar mortality patterns. Nevertheless, it could be expected that a male group presents higher mortality than a female group, due to genetic and biological reasons, and that individuals with higher educational level present lower mortality than lower educational level ones, due to socioeconomic factors \cite{preston2001demography}.

When considering the modelling of multi-populations using single population models methodologies, several problems may arise such as mortality curves crossing over at infant/young and oldest ages, where data are sparse or there is a decreasing mortality trend in age. In the English Life Tables (ELTs)  versions ELT14, ELT15 and ELT17, for males and females, these problems were addressed using rather ad hoc approaches to avoid divergence between mortality curves \citep{forster22}. The same issue occurs in the Brazilian insurance market life tables (BR-EMS) for male and female with death and survivorship coverage totalling four life tables. For example, in the BR-EMSv.2021, the female tables should close after the male ones and the survivorship coverage has higher longevity when compared to the death coverage \citep{BREMS2021}.

To successfully address this issue, \cite{forster22} propose a method for jointly smoothing male and female mortality rates by using adaptive splines. Their proposed model is able to produce smoothing for the entire age range and it is robust at the oldest ages where  data are usually limited.  Besides, joint modelling allows for borrowing information across sexes, avoiding divergence and intersection between mortality curves.
We extend the idea presented in \cite{forster22} by using dynamic linear models \citep[DLM,][] {West97} to smooth mortality rates. This class of models is usually applied in the context of time series analysis, to address intrinsically autocorrelated observations gathered through time $t=1,2,\ldots$.  We adopt a particular specification of the DLM class to produce graduated mortality tables, formally recognising the association between mortality rates for neighbouring ages (or age groups) and imposing smoothness of the estimated mortality curve in the transition between ages \citep{BayesMortalityPlus, silva2023}. The univariate mortality dynamic linear model is available on \texttt{BayesMortalityPlus} package in \texttt{R} (see \url{https://cran.r-project.org/package=BayesMortalityPlus}).

Using the well-known Kalman Filter proposed by \cite{Kalman} as a smoother method to obtain means and covariances of the conditional distributions of the evolution state equations does not account for all the useful information that the neighbours of age $x$ could provide. This has a trivial effect which is the incorrect estimation of uncertainty over transition age in the modelling. The Bayesian approach considered in this work performs estimation prospectively in the age domain but also retrospectively, which will allow all the available information to be used.  

In addition, another issue arises when information about mortality is missing or there are incomplete life tables for one or more populations. \cite{alexander2017flexible} considers a Bayesian hierarchical approach to account for correlation in the joint modelling of subnational mortality. In this scenario, a small population presents high variation leading to poor mortality graduation if a model for a single population is assumed due to the unclear pattern available for estimation. {\cite{li2020bayesian} propose a Gaussian state space version of the Lee-Carter model, adding features to address missing data, accounting for data from different sources, and reducing erratic parameter estimates due to data limitations. \cite{LiLee04} adapts the Lee-Carter method for forecasting mortality in populations with limited data. The authors introduce modifications to maintain accuracy with short time series, focussing on applications in countries with incomplete mortality records. The study demonstrates that the method remains robust even with sparse data. }

In mortality studies, pooling is fundamental for achieving more accurate and comprehensive analyses. Differences in mortality rates between genders contribute to understanding health trends, risk factors, and public health interventions, aiding policy planning, insurance calculations, and understanding population dynamics. Additionally, pooling addresses missing or incomplete data, a common challenge in mortality studies. By using more complete data from one gender to fill gaps in the other, pooling creates a more continuous and reliable picture of mortality trends, improving data quality and ensuring mortality estimates that better reflect true population dynamics.

The remainder of the text is structured as follows. {Section \ref{sec2} presents the structure of the multivariate data to be studied, based on data from the United Kingdom \citep{OFS2019}, and motivates the usefulness of pooling across subpopulations, emphasizing how shared information can improve inference and understanding across groups. Section \ref{sec3} introduces the proposed joint modelling of subpopulations, and the inclusion of a common term parameter, the age-varying smoothness parameter and the inference procedure.  Section \ref{sec4} includes methods to prevent curve crossover, convergence between subpopulations, and extrapolation at advanced ages. Section \ref{sec5} presents the distribution of the missing data.} Section \ref{sec6} discusses studies related to the behaviour of mortality tables at advanced ages, the use of age-varying smoothness parameters, joint modelling, and the evaluation of missing data. Section \ref{sec7} concludes with some final remarks. {Appendices \ref{apA} and \ref{apB} present details on inference procedure and model comparison criteria.}

\section{Pooled mortality models: motivation}\label{sec2}

\subsection{Mortality in England and Wales}

A significant challenge in mortality data analysis arises from the inherent sparsity of data at advanced ages. This issue is particularly pronounced when studying mortality patterns for populations in which certain age groups are underrepresented. In such cases, mortality estimates tend to become highly variable, and unrealistic behaviours such as the crossover of mortality curves between subpopulations may emerge. This crossover, where mortality rates for one group surpass those of another in unexpected age ranges, contradicts established demographic knowledge \cite{Jacques2001, preston2001demography}. 

\begin{figure}[H]
\centering
\includegraphics[width=0.6\textwidth]{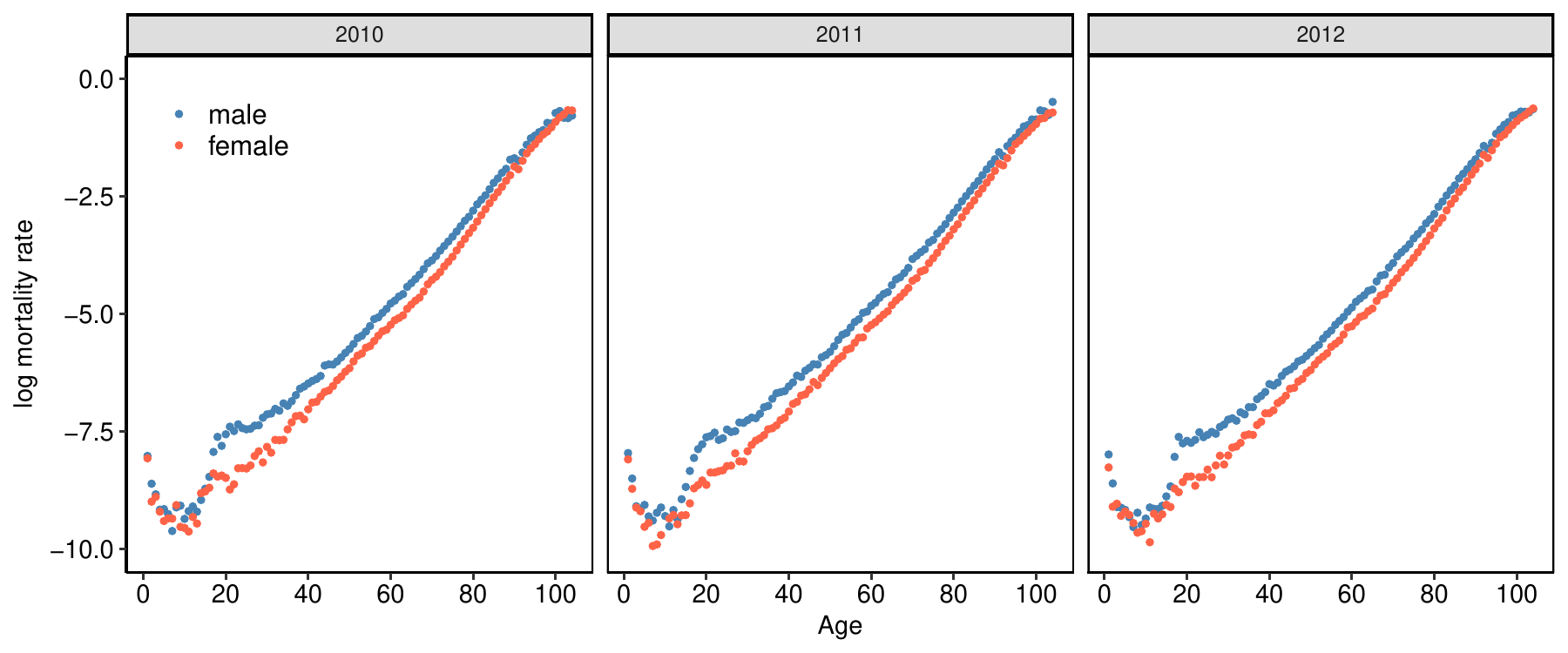}
\includegraphics[width=0.6\textwidth]{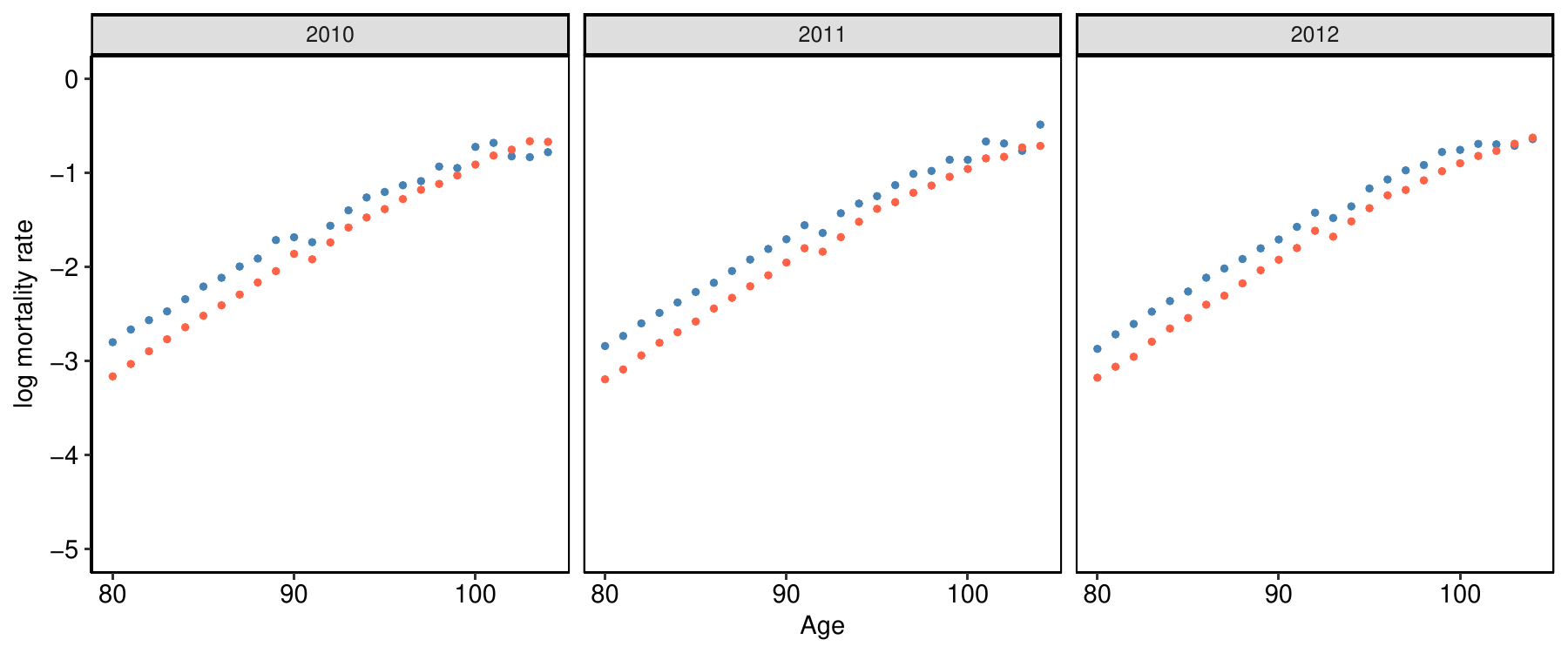}
\includegraphics[width=0.6\textwidth]{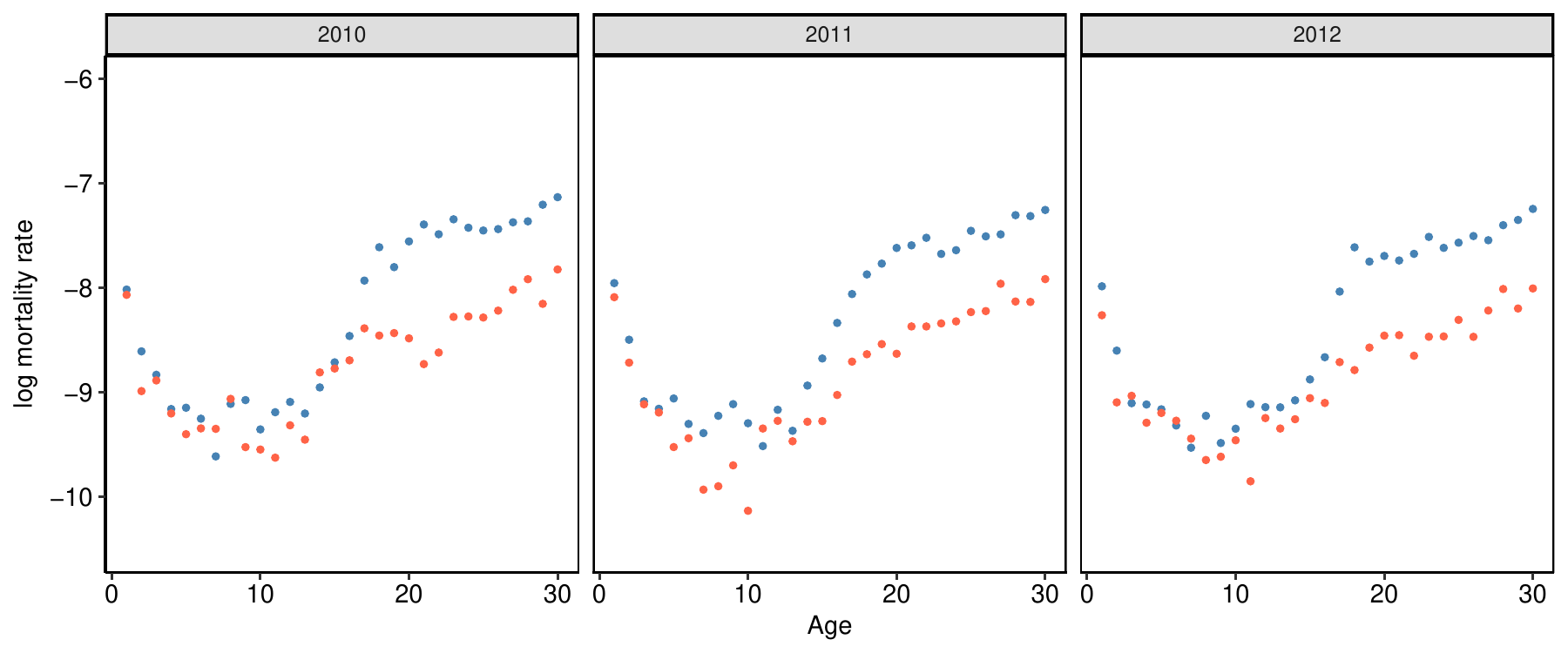}
\caption{Raw mortality rates in log-scale for England and Wales (E+W) from 2010 to 2012 for ages 1-104. The second and third rows zoom in on ages 80+ and 30-, respectively. Blue and red dots represent male ($j=1$) and female ($j=2$) populations.}
\label{fig:fig1}
\end{figure}

The aforementioned characteristics are observed in the mortality data analysed by \cite{forster22} for England and Wales (E+W) from 2010 to 2012,  which includes death counts and mid-year population estimates for males and females up to age 104, with individuals aged 105 and above grouped together. As shown in the panels of Figure \ref{fig:fig1}, the raw mortality rates on a log scale reveal typical age-specific trends: a decline in mortality during childhood; and a rise due to accidents, from late teenage to young age years, followed by a steady increase with age. However, data sparsity, especially at older ages, leads to increased dispersion and unreliable estimates, particularly for males in 2010.  

Modelling mortality at these advanced ages without proper smoothing can lead to undesirable extrapolations and discrepancies between the mortality rates of males and females. This is particularly problematic when trying to model and predict trends over time or across subpopulations. In response to these challenges, our approach integrates data from multiple subpopulations, in this case, namely, mortality from males and females, through a pooling strategy. This strategy allows us to borrow strength from similar populations and ensures that estimates are not unduly influenced by the limited data from specific subgroups. In addition, {we apply a 
sensitivity analysis on  age-varying smoothness as well as on prior specifications} {to avoid crossover between mortality curves}. {The joint model accounts for dependencies between subpopulations,} improving the overall accuracy and reliability of the estimates. By pooling information across subgroups, our model avoids the pitfalls of sparse data and provides more coherent and realistic mortality projections.

\subsection{Pooling across subpopulations}



In subnational mortality studies, some populations — particularly smaller regions or vulnerable groups — may suffer from extremely limited data, especially across certain age ranges. Analyzing these mortality curves independently can lead to implausible estimates, such as erratic patterns or unexpected crossovers (e.g., male mortality falling below female mortality at advanced ages). However, different subpopulations (e.g., by sex or socioeconomic status) often share common underlying mortality trends. Joint modeling across these groups facilitates the pooling of information, which not only stabilizes estimates in data-sparse contexts but also ensures greater demographic consistency and interpretability across populations.

\begin{figure}[H]
\centering
\begin{tabular}{cc}
\includegraphics[width=0.45\textwidth]{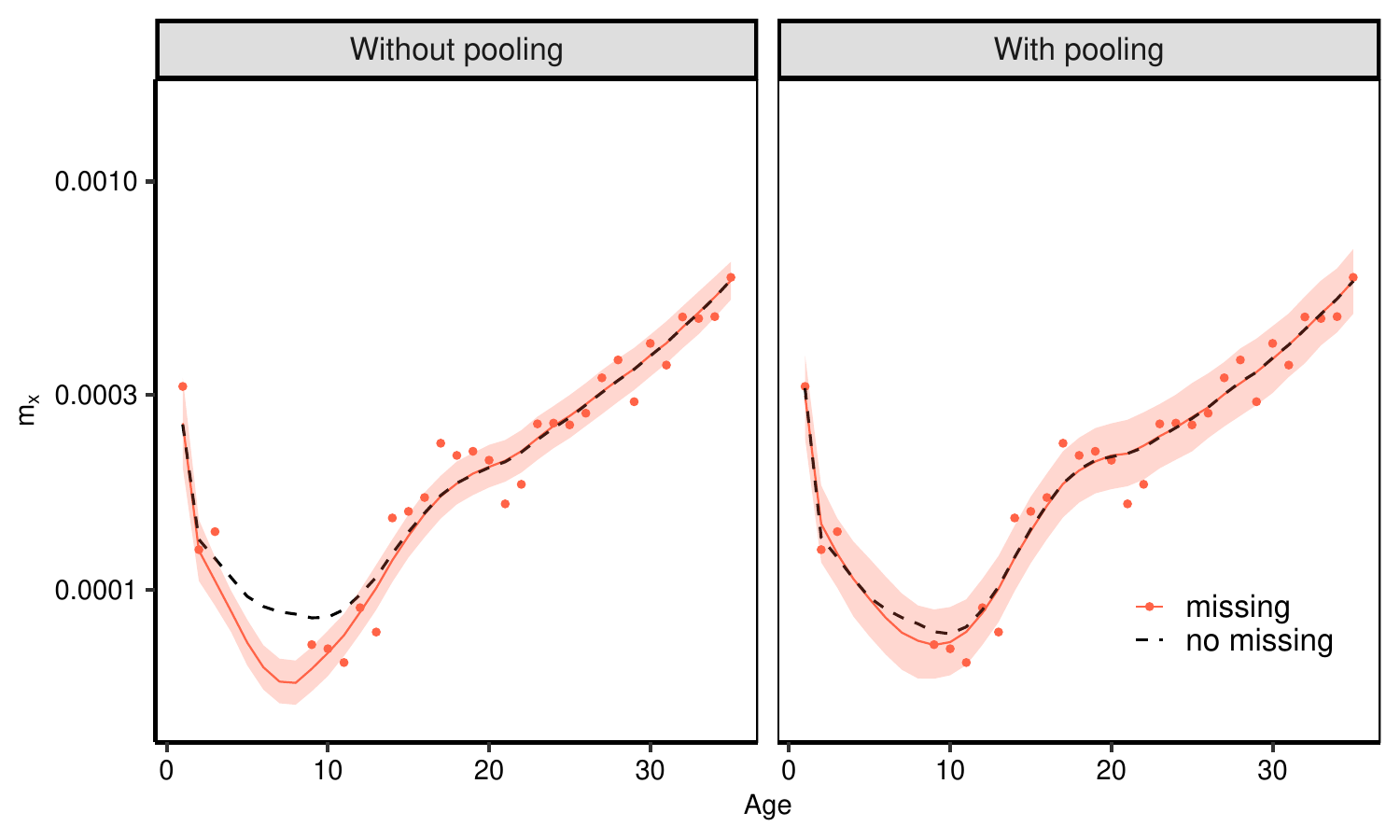} &  \includegraphics[width=0.45\textwidth]{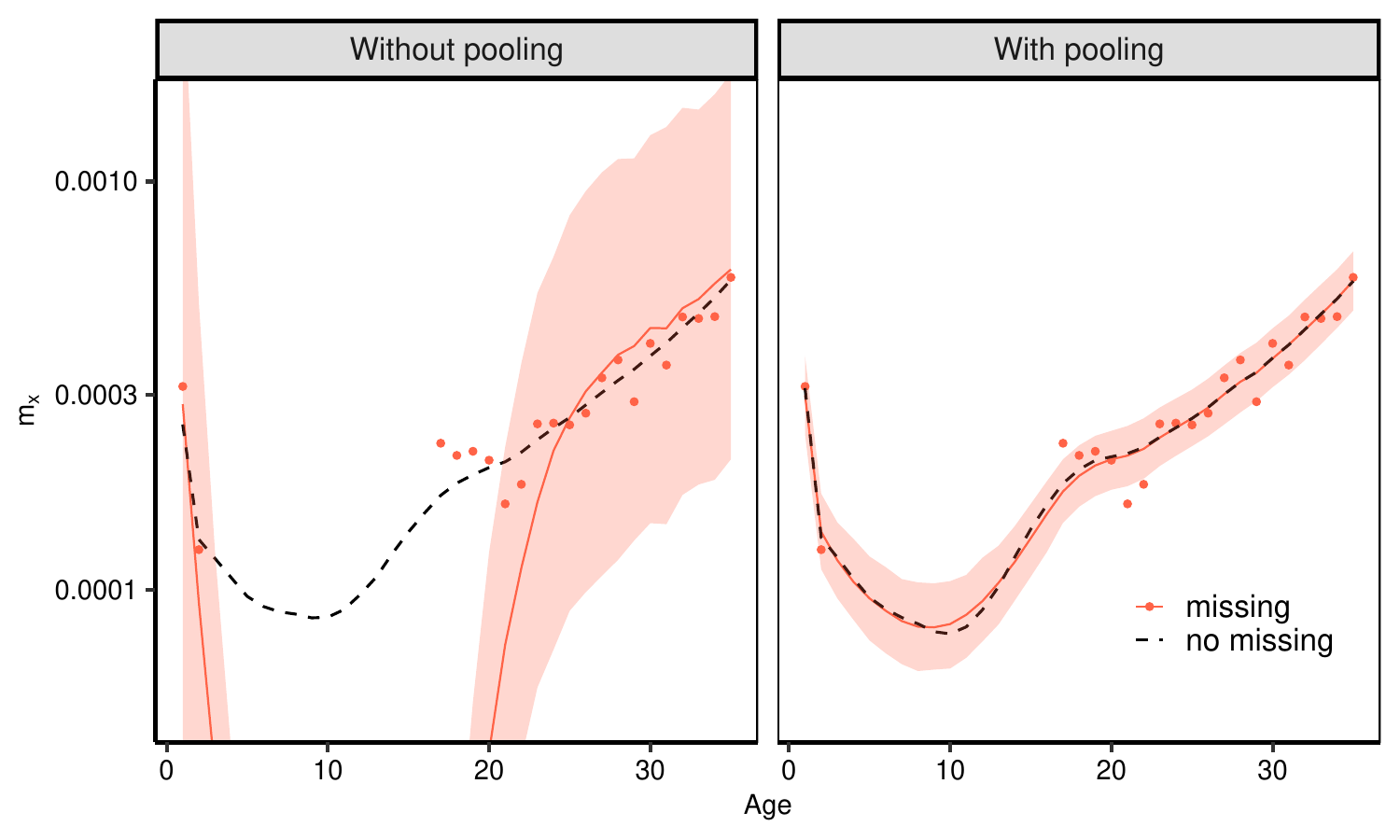}\\
5\% missing data  & 15\% missing  data \\
\end{tabular}
    \caption{Female mortality fits with 95\% predictive credible intervals without pooling and with pooling, considering 5\% (left panel) and 15\% (right panel) of missing data. Red dots represent the observed mortality data on the original scale. Red solid lines show the fit obtained using the incomplete data. Black dashed lines represent the hypothetical scenario in which the fit was obtained using the complete dataset.}
\label{sec2fig:fig1}
\end{figure}

By incorporating pooling, information is effectively borrowed across related groups, leading to smoother, more stable mortality estimates. This approach mitigates erratic behaviour in data-scarce age intervals, ensuring that mortality curves align with demographic expectations. {Figure \ref{sec2fig:fig1} (zoomed in 1-40 ages) illustrates this effect, comparing {female} mortality estimates generated with and without pooling { for two scenarios:} 5\% and 15\% of missing data, respectively in the original scale.} {In the scenario with 5\% missing data, the coupled estimation process used information from the male mortality curve for the same population and period as the target female table. In the scenario with a higher proportion (15\%) of missing data, additional information was necessary and the table with incomplete female data was jointly fit with a male  mortality table as well as the female  mortality curve itself, corresponding to an earlier period, thus resulting in the pooling of three mortality curves.}  The non-pooled estimates display unnatural fluctuations and crossover, while the pooled estimates maintain a coherent and demographically plausible structure. For comparison, the plot also displays lines representing a hypothetical scenario in which the fit was obtained using the complete dataset.

As follows, we present a methodology based on the polling approach which is able to perform well when data is scarce. 

\section{The multivariate dynamical linear mortality model}\label{sec3}

In this section, we outline our proposal for smoothing mortality tables based on dynamic linear models \citep{West97}.  The model consists of an observational equation that specifies the probabilistic distribution of the response variable, with a dynamic mean governed by a local linear predictor driven by latent dynamic states. At a higher level, a system equation describes the stochastic evolution of these latent states usually indexed in time, introducing temporal dependence and correlation among observations. In this proposal, the stochastic evolution is defined along the age index, inducing correlation across neighboring ages and enabling smooth transitions in mortality rates. Polynomial trend models are a specific application of this framework, offering flexibility to capture linear or non-linear trend patterns with varying slopes and concavities across the fitting interval. We adopt a second-order polynomial trend model to generate graduated mortality tables. This model can handle non-linear patterns commonly seen in log-mortality curves across different ages. 

\subsection{Dynamic linear graduation model}\label{GE}

{The central mortality rate for population $j$ and the mid-year age $x$ will henceforth be referred to simply as  mortality rate and is defined as
\begin{equation}
    m_x^{(j)} = \frac{D_x^{(j)}}{E_x^{(j)}},\;j=1,\ldots,J,
\end{equation}
with $D_x^{(j)}$ denoting the number of deaths and $E_x^{(j)}$, the population exposed to risk, at age $x$, for population $j$. Let $Y^{(j)}_x=\log \left(m_x^{(j)}\right)$ be the log death rate for the {\it j-th} population at age $x$. We consider the following model: 
\begin{eqnarray}
Y^{(j)}_x &=& \mu^{(j)}_x + v^{(j)}_x,~~~~v^{(j)}_x \sim \mathcal{N}(0,V^{(j)}) \label{eq.obs1}\\
\mu^{(j)}_x &=& \mu^{(j)}_{x-1} + \beta^{(j)}_{x-1} + w^{(j)}_{x,1},~~~~w^{(j)}_{x,1} \sim \mathcal{N}(0, W_{x,1}^{(j)}) \label{eq.evol.mu1}\\
\beta^{(j)}_x &=& \beta^{(j)}_{x-1} + w^{(j)}_{x,2},~~~~w^{(j)}_{x,2} \sim \mathcal{N}(0, W_{x,2}^{(j)}), \label{eq.evol.beta1}
\end{eqnarray}
\noindent where $x=x_1, \ldots, \vartheta$,  and $\vartheta$ define the upper limit for the age window considered for mortality graduation}. The random errors $v^{(j)}_x$, $w^{(j)}_{x,1}$ and $w^{(j)}_{x,2}$ are assumed sequentially independent following Gaussian distributions.
The state $\mu^{(j)}_x$ denotes the dynamic level of the log-mortality, with stochastic evolution between consecutive  ages guided by equation (\ref{eq.evol.mu1}), and $\beta^{(j)}_x$  denotes the local slope of the log-mortality curve at age $x$, allowing for different gradients through ages, since $\beta^{(j)}_x$ evolves according to the random walk described in equation (\ref{eq.evol.beta1}). The evolution structures in equations (\ref{eq.evol.mu1}) and (\ref{eq.evol.beta1}) formally accommodate the autocorrelation inherent in mortality patterns across ages.  Furthermore, we assume that the probability of death $q^{(j)}_x$ at age $x$ for the $j$-th population is determined by the relationship $q^{(j)}_x = 1 - \exp (- m^{(j)}_x) = 1 - \exp (- \exp(y^{(j)}_x))$, where $ q^{(j)}_x \in (0,1)$ for $j=1, \ldots, J$.

The model can be rewritten in the general form of a dynamic linear model {for log-mortality} as follows:
\begin{eqnarray}
{\bf Y}_x&=&{\bf F}_x \boldsymbol{\theta}_x+ {\bm v}_x, \quad \quad \thinspace \thinspace {\bm v}_x \sim \mathcal{N}_{J}({\bf 0},{\bf V}),  \label{eq:joint_obs}\\
\boldsymbol{\theta}_x &=& {\bf  G}_x\boldsymbol{\theta}_{x-1} + \boldsymbol{\omega}_x, \quad \boldsymbol{\omega}_x\sim \mathcal{N}_p({\bf 0},{\bf W}_x), \label{eq:joint_evo}
\end{eqnarray}

\noindent where ${\bm Y}_x= \left({Y}_x^{(1)}, \ldots,{ Y}_x^{(J)}\right)'$ is a vector of log mortality rates for 
$J$ populations, $\boldsymbol{\theta}_x= \left(\boldsymbol{\theta}_x^{(1)}, \ldots,\boldsymbol{\theta}_x^{(J)}\right)'$ is the state vector with $\boldsymbol{\theta}^{(j)}_{x}= (\mu^{(j)}_x,\beta^{(j)}_x)'$, $j=1, \ldots, J$.  {The observational errors  ${\bm v}_{x} = \left(v^{(1)}_{x}, \ldots, v^{(J)}_{x} \right)'$ follow a $\mathcal{N}_{J}\left({\bm 0}, {\bf V }\right)$ distribution and the system errors ${\bf w}_{x}= \left({\bf w}^{(1)}_x, \ldots, {\bf w}^{(J)}_x \right )'$ follow a  $\mathcal{N}_{2J}\left({\bm 0}, {\bf W}_x\right)$, where ${\bf w}^{(j)}_x= (w^{(j)}_{x,1},w^{(j)}_{x,2})'$}. For each population $j$:

\begin{equation*}
{\bf G}^{(j)}_x = 
\begin{bmatrix}
1 & 1 \\ 0 & 1
\end{bmatrix},
~~~~{\bf W}^{(j)}_x = 
\begin{bmatrix}
W_{x,1}^{(j)} & W_{x,3}^{(j)} \\ W_{x,3}^{(j)} & W_{x,2}^{(j)}
\end{bmatrix},
~~~~{\bf F}^{'(j)}_x = 
\begin{bmatrix}
1 & 0
\end{bmatrix},
\end{equation*}

\noindent where \({\bf W}_{x}^{(j)}\) represents the states covariance matrix, \({\bf G}_{x}^{(j)}\) is the evolution matrix, and \({\bf F}_{x}^{(j)}\) is the observation matrix. The evolution matrix \({\bf G}_{x}^{(j)}\) drives the relationship between the current  and previous states, indicating how each state variable evolves over ages. The state covariance matrix \({\bf W}_{x}^{(j)}\) quantifies the uncertainty associated with the states, while the observation matrix \({\bf F}_{x}^{(j)}\) links the state vector and the mean log-mortality. The components of the joint model (\ref{eq:joint_obs})-(\ref{eq:joint_evo}) for the $J$ populations are obtained by superposition of individual population components, as follows:

\begin{equation*}
{\bf F}_x = 
\begin{bmatrix}
{\bf F}^{'(1)}_x & \bm{0} & \cdots & \bm{0} \\
\bm{0} & {\bf F}^{'(2)}_x & \cdots & \bm{0} \\
\vdots & \vdots & \ddots & \vdots \\
\bm{0} & \bm{0} & \cdots & { \bf F}^{'(J)}_x
\end{bmatrix},
~~~{\bf V} = \begin{bmatrix}
V^{(1)} & \sigma_{12}^2 & \cdots & \sigma_{1J}^2 \\
\sigma_{21}^2 & V^{(2)} & \cdots & \sigma_{2J}^2 \\
\vdots & \vdots & \ddots & \vdots \\
\sigma_{J1}^2 & \sigma_{J2}^2 & \cdots & V^{(J)}
\end{bmatrix},  
\end{equation*}
\vspace{0.3cm}
\begin{equation*}
{\bf G}_x = 
\begin{bmatrix}
 {\bf G}^{(1)}_x & \bm 0 & \cdots & \bm 0 \\
\bm 0 & {\bf  G}^{(2)}_x & \cdots & \bm 0 \\
\vdots & \vdots & \ddots & \vdots \\
\bm 0 & \bm 0 & \cdots & {\bf G}^{(J)}_x
\end{bmatrix},
~~~{\bf W}_x = 
\begin{bmatrix}
{\bf  W}^{(1)}_x & \bm{0} & \cdots & \bm{0} \\
\bm{0} & {\bf  W}^{(2)}_x & \cdots & \bm{0} \\
\vdots & \vdots & \ddots & \vdots \\
\bm{0} & \bm{0} & \cdots & {\bf    W}^{(J)}_x
\end{bmatrix}. \\
\end{equation*}
\vspace{.2cm}

The Markovian structure in equation (\ref{eq:joint_evo}) induces the association between different age states, enabling us to capture complex relationships and trends, along ages. The model assumes that log-mortalities across different ages are correlated through the shared underlying state vector \(\boldsymbol{\theta}_x\). This correlation facilitates better estimation of parameters and more accurate predictions. Details on the estimation of the latent components in the model are discussed in Appendix \ref{apA}. For the specific case of joint modelling of $J=2$ populations, the observational can be written as:
\begin{align*}
Y^{(1)}_x &= \mu^{(1)}_x  + v^{(1)}_x \\
Y^{(2)}_x &= \mu^{(2)}_x + v^{(2)}_x, 
\end{align*}
and evolution of state equations are specified as
\begin{align*}
~~~ \mu^{(1)}_x &= \mu^{(1)}_{x-1} + \beta^{(1)}_{x-1} + w^{(1)}_{x,1} \\
\beta^{(1)}_x &= \beta^{(1)}_{x-1} + w^{(1)}_{x,2} \\
\mu^{(2)}_x &= \mu^{(2)}_{x-1} + \beta^{(2)}_{x-1} + w^{(2)}_{x,1} \\
\beta^{(2)}_x &= \beta^{(2)}_{x-1} + w^{(2)}_{x,2},
\end{align*}
\noindent  which corresponds to the form of equations (\ref{eq:joint_obs}) and (\ref{eq:joint_evo}), with:
$$
\mathbf{F}_x = \begin{pmatrix}
1 & 0 & 0 & 0 \\
0 & 0 & 1 & 0 \\
\end{pmatrix},
~~~\mathbf{G}_{x} = 
\begin{pmatrix}
1 & 1 & 0 & 0 \\
0 & 1 & 0 & 0 \\
0 & 0 & 1 & 1 \\
0 & 0 & 0 & 1 \\
\end{pmatrix},
~~~ \boldsymbol{\theta}_x =
 \begin{pmatrix}
 \mu^{(1)}_x \\
 \beta^{(1)}_x \\
 \mu^{(2)}_x \\
 \beta^{(2)}_x
 \end{pmatrix},
~~~{\bf V} = 
\begin{pmatrix}
V^{(1)} & \sigma^2_{12} \\
\sigma^2_{21} & V^{(2)} \\
\end{pmatrix},
$$
where ${\bf F}_x= blockdiag \left({\bf F}_x^{(1)},{\bf F}_x^{(2)}\right)$  and  ${\bf G}_x = blockdiag \left({\bf G}_x^{(1)},{\bf G}_x^{(2)}\right)$. \\

\subsection{Modelling a common component across subpopulations} 

{In the formulation presented so far, the covariances $\sigma^2_{ij}$, $i,j = 1, \ldots, J$, are the elements that explicitly reflect associations between the log-mortalities of the different populations under analysis. We additionally introduce a common term $\alpha_x$ in the log-mortality model, facilitating the explicit sharing of information between populations, with clear interpretation}. This common parameter acts as a baseline mortality rate that evolves with age and is adjusted according to specific variations observed in each population. This approach smooths out fluctuations in mortality rates, resulting in more accurate and robust estimates and providing a cohesive understanding of mortality dynamics across various groups.

Additionally, the common term $\alpha_x$ is particularly useful for populations with missing data. It allows the model to leverage information from populations with more complete data to fit mortality curves for associated populations with missing data, leading to more reliable predictions. By identifying and incorporating common patterns and trends in mortality,  model robustness and precision are enhanced, leading to more accurate estimates. Thus, we reparameterise the multivariate dynamic linear model given by equations (\ref{eq.obs1}), (\ref{eq.evol.mu1}) and (\ref{eq.evol.beta1}), including the parameter $\alpha_x${, and $\mu_x^{(j)}$ now denotes the deviation in the log-mortality level of population $j$ at age $x$ relative to the global level $\alpha_x$,}  as follows:

\begin{eqnarray}
Y^{(j)}_x &=& {\alpha_x} + \mu^{(j)}_x + v^{(j)}_x,~~~~~~~~~v^{(j)}_x \sim \mathcal{N}(0,V^{(j)}) \label{eq.obs}\\
\mu^{(j)}_x &=& \mu^{(j)}_{x-1} + \beta^{(j)}_{x-1} + w^{(j)}_{x,1},~~~~w^{(j)}_{x,1} \sim \mathcal{N}(0,W^{(j)}_{x,1}) \label{eq.evol.mu}\\
\beta^{(j)}_x &=& \beta^{(j)}_{x-1} + w^{(j)}_{x,2},~~~~~~~~~~~~~~~w^{(j)}_{x,2} \sim \mathcal{N}(0,W^{(j)}_{x,2}), \label{eq.evol.beta}\\
{\alpha_x} &{=}& {\alpha_{x-1} + \sum_{j=1}^{J} \mu_{x-1}^{(j)} + w^{*}_x, ~~~~w^{*}_{x} \sim \mathcal{N}(0,W^{*}_x), \label{eq.c}} 
\end{eqnarray}
\noindent where $j=1, \ldots, J$ represents the populations and $x=x_1, \ldots, \vartheta$, with $x_1$ being the minimum age and $\vartheta$ the maximum age. Notice that $\alpha_x$ depends on the dynamic levels of all mortality curves, capturing shared trends across populations for improved analysis and predictions.
Again, assuming we have male ($j=1$) and female ($j=2)$ groups, the model can be written using the specification through of the quadruple \{\( \mathbf{F}_x \), \( \mathbf{G}_x \),  $\mathbf{V}$, $\mathbf{W}_x$\} as:

$$
\mathbf{F}_x = \begin{pmatrix}
1 & 0 & 0 & 0 & 1 \\
0 & 0 & 1 & 0 & 1
\end{pmatrix},
~~ \mathbf{G}_x = \begin{pmatrix}
1 & 1 &0 & 0 & 0 \\
0 & 1 & 0 & 0 & 0 \\
0 & 0 & 1 & 1 & 0 \\
0 & 0 & 0 & 1 & 0 \\
1 & 0 & 1 & 0 & 1
\end{pmatrix},
~~\boldsymbol{\theta}_x =
\begin{pmatrix}
\mu^{(1)}_x \\
\beta^{(1)}_x \\
\mu^{(2)}_x \\
\beta^{(2)}_x \\
\alpha_x \\
\end{pmatrix},
~~{\bf V} = 
\begin{pmatrix}
V^{(1)} & \sigma^{2}_{12} \\
\sigma^{2}_{21} & V^{(2)} \\
\end{pmatrix}.
$$

Notice that the flexibility of the dynamic linear model is imposed by the specifications of the matrix {\bf F} and {\bf G}. The parameters $\boldsymbol{\theta}_x$ and ${\bf V}_x$ are estimated using the Forward Filtering and Backward Sampling (FFBS) algorithm (\cite{Sylvia1994} and \cite{Carter1994}), which effectively captures the dynamics of the underlying states. The model is completed with subject information about the latent states at age $x_0$ (an age immediately inferior to those for which mortality data are available). This specification can be vague and expressed as $\bm \theta_{x_0} \sim N_p (\bm m_{x_0}, \bm C_{x_0})$, with $\bm m_{x_0}, \bm C_{x_0}$ chosen by the analyst. For instance, $\bm C_{x_0}$ can be specified as a diagonal matrix with large variances,  reflecting no prior knowledge about the latent states. {On the other hand, the state covariance matrix ${\bf W}_x$ is implicitly defined via a discount factor approach, to be detailed in the following sections, which calibrates the influence of past-age mortality on the state estimates for each age $x$. Appendix \ref{apA} shows the detailed posterior distribution and inference procedure for the general model.} 

\subsection{Age-Varying Smoothness}\label{sec3.2}

{The degree of smoothness achieved in the log-mortality fit, across ages, can be controlled by applying discount factors \citep[][Chap.6]{West97} to specify the covariance structure of the evolution errors  $\boldsymbol{\omega}_x$ in equation (\ref{eq:joint_evo}).} These factors influence how quickly the model adapts to new observations. This property is of particular interest in our proposal, as it will allow for smoothness for nearby ages. The discounting strategy introduces a form of regularization that smooths the fitted log-mortality curve by weighting past age information differently from current observations. This allows the model to gradually adjust to new data without overreacting to random fluctuations. Specifically, the factor is bounded between 0 and 1, with values close to 1 implying more emphasis on past data, resulting in smoother transitions and slower adaptation to changes. On the other hand, a lower discount factor allows for quicker adaptation but at the cost of more sensitivity to short-term fluctuations.

{This behaviour is induced by defining the covariance of the state evolution equation as  
\begin{equation}
{\bf W}_x= \frac{1-\delta}{\delta} {\bf P}_x, \label{W.desc}
\end{equation}  
where \( {\bf P}_x = {\bf G}_x {\bf C}_{x-1} {\bf G}_x' = \text{Cov}[{\bf G}_x \boldsymbol{\theta}_{x-1} \mid D_{x_0}, {\bf y}_{x_1: x-1}] \). Here, \( D_{x_0} \) denotes any initial subjective information on the log-mortality at early ages, and \( {\bf y}_{x_1: x-1} \) represents the observed log-mortality rates up to age \( x-1 \). It follows from the evolution equation~(\ref{eq:joint_evo}) that \( {\bf W}_x \) acts as an inflation factor applied to \( {\bf P}_x \), as the model transitions from posterior inference at age \( x-1 \) to prior inference at age \( x \).} For a detailed description see Appendix \ref{apA}.

{For each age $x$,} the choice of  $\delta_x$ determines the smoothness of the log-mortality curve {around $x$}. When $\delta_x \approx 1$, the model produces smoother trajectories with near-zero evolution variances for level and slope, {implying that, except for a deterministic transformation ${\bf G}_x$,  the latent states $\boldsymbol{\theta}_x$ are similar to $ \boldsymbol{\theta}_{x-1}$. Conversely, values further from 1 allow the model to adapt more quickly to the observed mortality at age $x$. In the extreme case where $ \delta_x = 1$, the model becomes linear with constant intercept and slope across all ages. Sensitivity analysis can be performed by fitting models with different $ \delta_x $ values and using model selection metrics. Additionally, different $\delta_x$ values can be applied across age intervals to accommodate varying levels of mortality rate changes, such as using values close to 1 for stable age intervals and deviating from 1 for more variable intervals. }

\begin{figure}[H]
    \centering
    \includegraphics[width=0.6\linewidth]{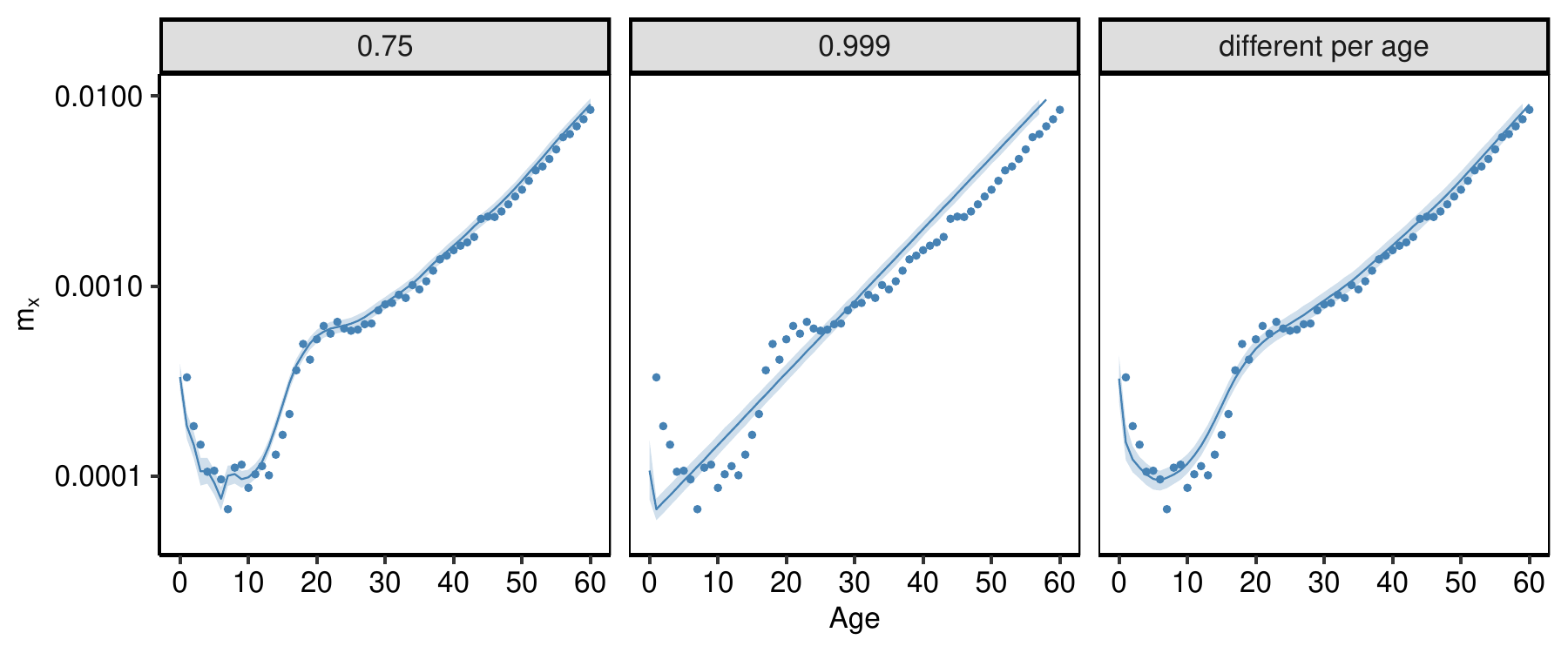}
    \caption{Illustration shows the flexibility of discount factors considering England and Wales, 2010 male with range age 1-60, via \texttt{BayesMortalityPlus} \texttt{R} package.}
    \label{fig:figagevarying}
\end{figure}

  {To illustrate the flexibility of using discount factors, Figure \ref{fig:figagevarying} (zoomed in on ages 1 to 60) shows the univariate dynamic linear fit for the male population in the year 2010 and age interval 1-104, considering three scenarios: fixed discount factors \(\delta = 0.75\) and \(\delta = 0.9999\) applied uniformly across all ages, as well as varying discount factors by age ranges, \(\delta_x\), where \(\delta_{1,1:5} = 0.99\), \(\delta_{2,6:35} = 0.80\), \(\delta_{3,36:85} = 0.85\), and \(\delta_{4,86+} = 0.99\), covering the age groups of 1 to 5, 6 to 35, 36 to 85 years, and 85+, respectively. The fit was performed using the \texttt{BayesMortalityPlus} package available in R (\cite{silva2023}, \cite{BayesMortalityPlus}).} In our proposal, flexibility is added to the model specification by allowing the discount factor to vary in different regions of the age domain and by subpopulations. In adult ages, mortality behaves linearly, and more dependency on the past is allowed. In contrast, for younger ages, where mortality rates may exhibit more abrupt changes due to varying health conditions or demographic shifts, age-varying smooth can be adjusted to a value further from 1 to accommodate these variations and avoid excessive smoothing. This will result in differential rates of change in the underlying mortality, enhancing the ability of the model to capture complex dynamics. Note, however, that proper choice of discount factors is crucial for curve estimation performance as too low factors can lead to overfitting, while too high a factor might result in the model failing to capture significant changes in the data. In the context of mortality graduation, three or four  factors for different age intervals are usually enough to capture the typical behaviours observed in the data. 

\section{Model for older ages, extrapolation and convergence between subpopulations}\label{sec4}

This section focuses on the prediction of mortality trends at older ages. We first present an approach based on dynamic linear models (DLM), which allows for mortality extrapolation while accommodating different assumptions regarding longevity limits. Additionally, we examine the phenomenon of convergence in mortality trends across subpopulations. Finally, we introduce a methodology based on prior and smoothness specifications to prevent cross-over between mortality curves of different subpopulations, ensuring coherence in demographic analyses and long-term forecasting.

\subsection{Extrapolating using DLM method}

{Fitting mortality models for advanced ages presents significant challenges due to the decrease in data quality for older age groups. Two primary approaches are typically considered: (1) employing a distinct model specifically tailored for the older age domain \citep{Forster2018}, or (2) extrapolating mortality rates by making predictions using models fitted to younger age groups. In this study, we adopt the second approach, utilising extrapolations from our proposed dynamic linear model. Specifically, we explore various scenarios to account for differing hypotheses regarding the potential upper limits of the human lifespan.

Let $\vartheta$ represent the threshold that defines the advanced ages $x > \vartheta$, from which extrapolation will be applied.} The predictive distribution for $k$ steps ahead that extrapolate the mortality curve fitted by our proposed model can be derived as follows, according to \cite{West97}. For $k = 1, 2, \ldots $, the predictive distribution is obtained through the following process: 
\begin{align}
\pi({\bm \theta}_{\vartheta+k}\mid {\bf y}_{x_1:\vartheta}) &= \int \pi({\bm \theta}_{\vartheta+k}|{\bm \theta}_{\vartheta+k-1}) \pi({ \bm\theta}_{\vartheta+k-1}\mid {\bf y}_{x_1:\vartheta}) \: d{\bm \theta}_{\vartheta+k-1} \\
f({\bf y}_{\vartheta+k} \mid {\bf y}_{x_1:\vartheta}) &= \int f({\bf y}_{\vartheta+k} \mid {\bm \theta}_{\vartheta+k}) \pi({\bm \theta}_{\vartheta+k} \mid {\bf y}_{x_1:\vartheta}) \: d{\bm \theta}_{\vartheta+k},\nonumber
\end{align}
\noindent where \( f(\cdot) \) denotes the observational model conditional on the latent states \( {\bm \theta} \), \( \pi({\bm \theta}_{\vartheta+k}|{\bf y}_{x_1:\vartheta}) \) denotes the prior distribution for \( {\bm \theta} \), \( k \) steps ahead, and \( f({\bf y}_{\vartheta+k} \mid {\bf y}_{x_1:\vartheta}) \) is the predictive distribution for the observation (log-mortality rate) \( k \) steps ahead. Considering extrapolation for \( k \) ages ahead, we obtain the forecast distributions for \( h \) steps ahead, where \( h = 1, 2, \ldots, k \), conditional on the information up to the maximum age $\vartheta$ used in the model fitting. 
{This result is obtained through simulation using the FFBS algorithm, as detailed in Appendix \ref{apA}.

{An essential question in this context is the plausibility of the terminal ages predicted by our model. Unconstrained extrapolations may yield unrealistic lifespan projections. Therefore, we examine scenarios, as commonly done in time series analysis, to set reasonable boundaries. At one extreme, we apply our general model with flexible assumptions, while at the other extreme, we assume a restricted subclass of the model in which a single curve is fitted across subpopulations to impose greater consistency.}

{This framework is flexible to accommodate a wide range of behaviours. In particular, it allows for convergence of mortality curves among subpopulations at older ages, a phenomenon frequently observed, particularly when mortality rates are extrapolated across advanced age groups. For example, in the context of sex differences, females have historically shown higher life expectancy and distinct mortality patterns compared to males. However, these differences tend to diminish with advancing age, as mortality risk factors become more similar across populations. Thus, it is of interest to evaluate this convergence in the joint modelling process of the curves. For this, we consider the more flexible scenario, with the unconstrained model at the beginning of extrapolation and mix this scenario with the scenario that assumes a single curve for the subpopulations at the end of our lifespan. Selecting a realistic terminal age for the proposed table will help determine the point in time at which the two scenarios should be combined.}

\subsection{Preventing cross-over between populations }

Although the previous model considers joint modelling of subpopulations, the crossing of mortality curves is not guaranteed. In several contexts preventing cross-over in mortality rates between subpopulations is a crucial aspect of mortality analysis. Cross-over can occur not only at older ages, where the mortality rates of one subpopulation might exceed those of another but also at younger ages, where data can be noisy and lead to misinterpretations. Therefore, in contexts where curve crossing is theoretically implausible due to demographic or domain-specific knowledge, it is crucial to ensure that mortality curves remain non-crossing across all age ranges to prevent erroneous interpretations of mortality trends.

The prevention of cross-over can be effectively achieved through a careful sensitivity analysis of the prior choice for the state parameter $\boldsymbol{\theta}_{x_0}$, combined with age-varying smoothness. The selection of hyperparameters for the prior specification of $\bm \theta_{x_0}$ is fundamental to defining the flexibility and robustness of the model, particularly in younger ages. In this context, well-informed prior choices help to incorporate relevant demographic knowledge and stabilise the model when data is sparse. Specifying the prior for the initial state is not merely a matter of dealing with uncertainty, but rather a means to incorporate prior knowledge in a structured and transparent way. In our proposed model, the state parameter prior follows a multivariate Gaussian distribution, $\mathcal{N}_p(\bm m_{x_0}, \bm C_{x_0})$, where $\bm m_{x_0}$ represents the initial mean and $\bm C_{x_0}$ the covariance matrix. These choices reflect the degree of confidence in the initial conditions, and selecting these hyperparameters thoughtfully is essential to leverage prior information effectively. Sensitivity analyses allow us to assess how variations in prior assumptions impact model performance, thereby enhancing robustness and interpretability.

When there is significant uncertainty about the initial state, $\bm C_{x_0}$ can be specified as a diagonal matrix with large variances, allowing the model to learn $\bm \theta_{x_0}$ from the data. This noninformative prior approach ensures flexibility by avoiding restrictive assumptions. However, when prior knowledge is available, a more informative prior can guide the model and stabilise estimates. It is essential to evaluate how different choices for $\bm C_{x_0}$ impact the performance of the model, as restrictive priors can cause biased results, while vague priors can cause instability. For a comprehensive discussion on the choice of priors for hyperparameters in hierarchical models, \citealp[see][]{gelman2006prior}.

Prior distributions incorporate external subjective information that complements the data. In joint models for multiple populations, priors can encode expectations that the mortality level of one population consistently surpasses another, even when data variability and empirical imprecision suggest otherwise. This ability to formally integrate external knowledge makes the Bayesian approach especially useful for preventing implausible cross-over in mortality curves.

\section{Missing data}\label{sec5}

Modelling missing data presents unique challenges in understanding and predicting mortality rates across populations. Missing mortality data are usual for data sets in which mortality events are relatively rare compared to the total population, resulting in a high proportion of zero values. 

Consider the $(n+q)-$dimensional vector ${\bf Y}_x= \left( {\bf Y}_x^{(obs)},{\bf Y}_x^{(miss)} \right)$, with ${\bf Y}_x^{(obs)}$ and ${\bf Y}_x^{(miss)}$ representing, respectively, the $n$-th dimensional observed mortality rates and $q$-dimensional missing values of ${\bf Y}_x$ at each age $x= x_1, \ldots, \vartheta$. In this case, ${\bf Y}_x^{(obs)}= (Y_x^{(1,obs)}, \ldots, Y_x^{(J,obs)})'$ and ${\bf Y}_x^{(miss)}= (Y_x^{(1,miss)}, \ldots, Y_x^{(J,miss)})'$ represent the vectors of observed and missing in each group $j$, respectively. To obtain samples from the posterior predictive distribution $p({\bf Y}_x^{(obs)} \mid {\bf Y}_x^{(miss)})$, assume that $ {\bm \theta}_x^{(k)}, {\bf V}^{(k)}$, represent the $k$-th sample from the joint posterior distribution $p({\bm \theta}_x, {\bf V} \mid {\bf Y}_x^{(obs)})$. Thus, samples from ${\bf Y}_x^{(miss)} \mid {\bf Y}_x^{(obs)}, {\bm \theta}_x^{(k)}, {\bf V}^{(k)}$ can be trivially obtained, applying properties of the partition of multivariate Normal distributions and the model assumption in equations (\ref{eq.obs1}), (\ref{eq.evol.mu1}) and (\ref{eq.evol.beta1}):
\begin{equation}\label{eq:missing}
    {\bf Y}_x^{(miss)} \mid  {\bf Y}_x^{(obs)}, {\bm \theta}_x, {\bf V} \sim \mathcal{N}_{q}\left(
{\bf F}'_{x} {\bm \theta}^{(miss)}_x +V_{m,o} V^{-1}_{o,o}\left({\bf y}^{(obs)}_{x} -{\bf F}'_{x} {\bm \theta}^{(obs)}_x \right), V_{m,m} - V_{m,o}V^{-1}_{o,o}V_{o,m}\right),
\end{equation}
\noindent with, 
$$
\boldsymbol{\theta}_x = 
\begin{pmatrix}
  \boldsymbol{\theta}^{(obs)}_x \\
  \boldsymbol{\theta}^{(miss)}_x \\
\end{pmatrix},
~~~{{\bf V}} = 
\begin{pmatrix}
V_{o,o} & V_{o,m} \\
V_{m,o} & V_{m,m} \\
\end{pmatrix},
$$
\noindent where $V_{o,o}$ is the $n \times n$ covariance matrix representing the covariance between the observed data, $V_{o,m}$, is the $n \times q$ covariance matrix representing the covariance between the observed and missing data, $V_{m,o}$ is the $q \times n$ covariance matrix, and $V_{m,m}$ is the $q \times q$ covariance matrix representing the covariance between the missing data.

Here missing values are handled as latent variables and are iteratively imputed based on their complete conditional distribution. This method updates the missing values in each iteration of a Markov Chain Monte Carlo (MCMC) scheme \citep{gamerman}, thus improving the estimates of the parameters and enhancing the robustness of the analysis, even with incomplete data. The initial values are crucial for Gibbs sampling in the presence of missing data. We follow the procedure outlined below:\\

\noindent {\it Step 1:} Estimate each missing mortality rate by performing linear interpolation between non-missing values, thus generating an initial complete dataset. \\

\noindent {\it Step 2:} Generate new values for missing data based on the complete conditional distribution given by equation (\ref{eq:missing}), then update the model parameters using the now complete data set. \\

\noindent {\it Step 3:} Repeat the process until the distributions of the parameters and the imputations stabilise, indicating the convergence of the algorithm.

\section{Data analysis}\label{sec6}

We analyse the mortality rates of the UK male and female populations from 2010 to 2012 based on \cite{OFS2019}, the same data studied by \cite{forster22}, and consider three case studies: first,  males and females mortalities are modelled separately to assess the fit at advanced ages while varying the discount factor. This allows the assessment of the effectiveness of univariate modelling for each subpopulation, as well as identification of its limitations when comparing subpopulations. 

The second case study focuses on joint modelling of male and female populations, allowing for shared information between the sexes. Our goal is to prevent crossover of the mortality curves and promote convergence between them. This approach not only maintains coherence in the mortality trends but also reduces estimation uncertainty by pooling data from both populations.

The third case study addresses the challenge of missing data and evaluates model performance relative to varying levels of data scarcity. Our objective is to assess how accurately the proposed model predicts mortality rates under these conditions. By comparing different strategies for managing missing data, we aim to enhance the robustness and reliability of the model, for mortality estimation. As model comparison criteria, we consider measures based on mean square prediction error (MSPE), mean absolute prediction error (MAPE), the width of credible intervals (WCI), assessing the predictive performance of the competing models. A brief description is presented in Appendix \ref{apB}.

\subsection{Study 1:  Mortality extrapolation at advanced ages with age-varying smoothness}\label{sec4.1}

This study 
aims to apply the dynamic linear model to the UK dataset, focusing on evaluating model performance, especially at advanced ages.  As seen in Figure \ref{fig:fig1} (second row), and Figures \ref{fig:fig2} and \ref{fig:fig3} (blue and red dots), mortality data for older ages are sparse. We investigate the same two scenarios presented in \cite{forster22}: (a) fitting the model using the full dataset, which includes ages 1 to 104, and (b) fitting the model with the last four oldest age groups removed, restricting the data set to ages 1 to 100. This study considers modelling the male and female mortality rates by fitting separate univariate models. Our proposed modelling approach leverages the flexibility afforded by incorporating discount factors into dynamic linear models, allowing for age-specific smoothness control.
 We illustrate this flexibility by assuming fixed factors ($\delta_x=\delta$ = 0.80, 0.85, 0.90, 0.95) for all ages, as well as varying discount factors $\delta_x$ across different age ranges. 

Figures \ref{fig:fig2} and \ref{fig:fig3}  exhibit separate fits by gender, for the total population of England and Wales (E+W), with extrapolated curves extending to age 120, along with respective 95\% predictive intervals for both age range scenarios. These figures display several discount factors for males (blue lines) and females (red lines) for the years 2010 to 2012 (arranged over rows). Figures  \ref{fig:fig4} and \ref{fig:fig5}  provide a detailed comparison of mortality curves, by sex, for ages 80 to 120, considering scenarios (a) and (b), respectively. Notice that when using a single discount factor for the age range 1-104, the estimated probability of death for both males and females tends to decrease at older ages, which contrasts with the expected increase in mortality with age. This trend is less pronounced when the age range is restricted to 1-100, suggesting that the extrapolation may be unrealistic and associated with higher uncertainty. Conversely, increasing the single discount factor $\delta$ closer to 1 tends to produce smoother extrapolated curves, yielding more consistent results when comparing the restricted age range of 1-100. However, the fit at younger ages remains less robust across both age range scenarios, as exemplified in the panel of Figure \ref{fig:fig2} ($\delta=0.95$, male, 2010, age 5-35).

The lack of robustness in the mortality fit may stem from the use of a fixed discount factor to control global smoothness, leading to insufficient smoothing of mortality rates across ages, particularly at older ages. In the last panel of Figures \ref{fig:fig2} and \ref{fig:fig3}, we introduce discount factors varying by age groups, $\delta_x$, where $\delta_{1,1:5}= 0.99$, $\delta_{ 2,6:35}=0.80$, $\delta_{3,36:85}=0.85$ and $\delta_{4,86+}=0.99$, covering ages 1 to 5, 6 to 35, 36 to 85 years and 85+ respectively. Compared to other panels in these figures, we observe that the age-varying factor produces a smoother fit across the entire age domain. Furthermore, increasing the discount factor (making it  closer to 1) around age 85 leads to a more robust fit at advanced ages and reduces estimation uncertainty compared to using a single discount factor across all ages. This approach is advantageous, as it specifically targets the regions with low exposures, where reliability  declines. Notably, adopting an age-varying discount  $\delta_x$ eliminates the need for ad hoc selection or the removal of the oldest ages for extrapolation. By allowing smoothness to vary across the entire age range, the model facilitates realistic extrapolation at advanced ages, where mortality data is less precise.

Thus far, we have modelled male and female mortality rates separately using univariate models. Figures \ref{fig:fig4} and \ref{fig:fig5} compare the fitted models for both sexes, using different advanced age intervals in the fitting: 80–104 in Figure \ref{fig:fig4}  and 80–100 in Figure \ref{fig:fig5}, both with extrapolation up to age 120. As shown,  assuming separate models results in cross-over extrapolations at advanced ages, which is counter-intuitive when comparing the mortality probabilities of these two subpopulations. {Moreover, cross-overs may also occur in the infant age groups.} Additionally, convergence between curves may take a considerable amount of time (e.g., in the scenario using the age range 1-100 with $\delta=0.95$). We expect male mortality to remain higher than female mortality, with both mortality curves converging at advanced ages, reflecting similar mortality probability behaviours in these subpopulations over time. To achieve convergence of male and female mortality curves when modelling them separately, it is typically necessary to apply specific constraints to ensure alignment at advanced ages. These constraints may involve selecting the age at which the convergence of mortality rates should begin and utilising transition functions that blend the male and female mortality curves from the chosen age onwards. Such functions can be tailored to smooth the transition and enforce the convergence of the curves.

\begin{figure}[H]
    \centering
    \begin{tabular}{c}
    \includegraphics[width=1\textwidth]{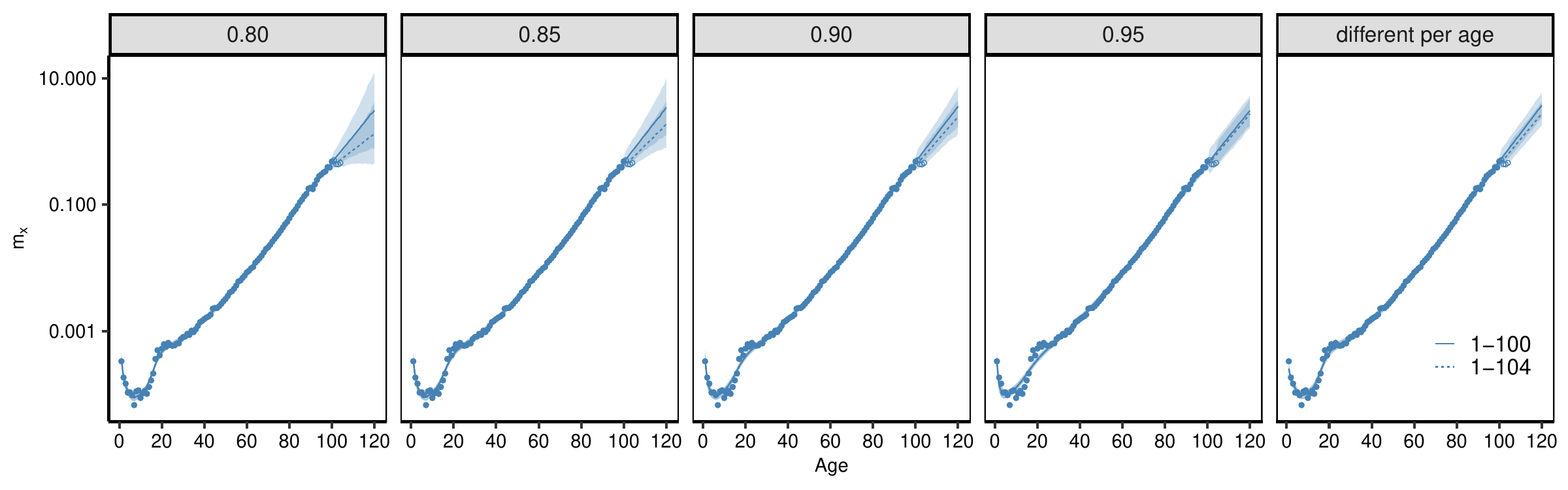}\\
        \includegraphics[width=1\textwidth]{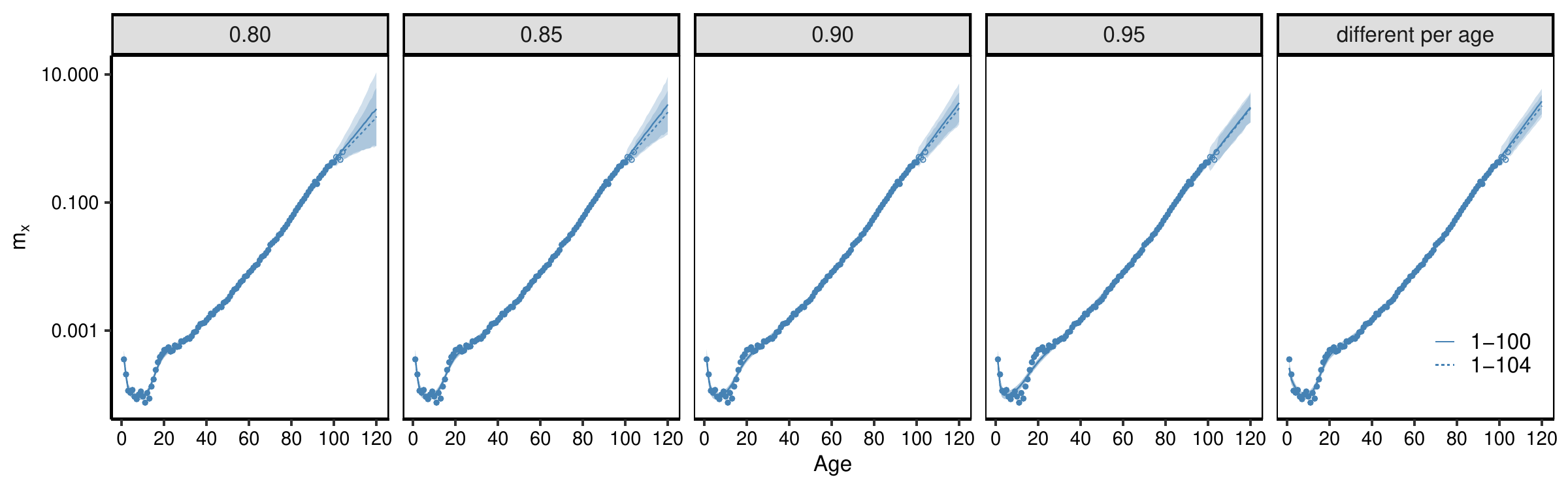}\\
            \includegraphics[width=1\textwidth]{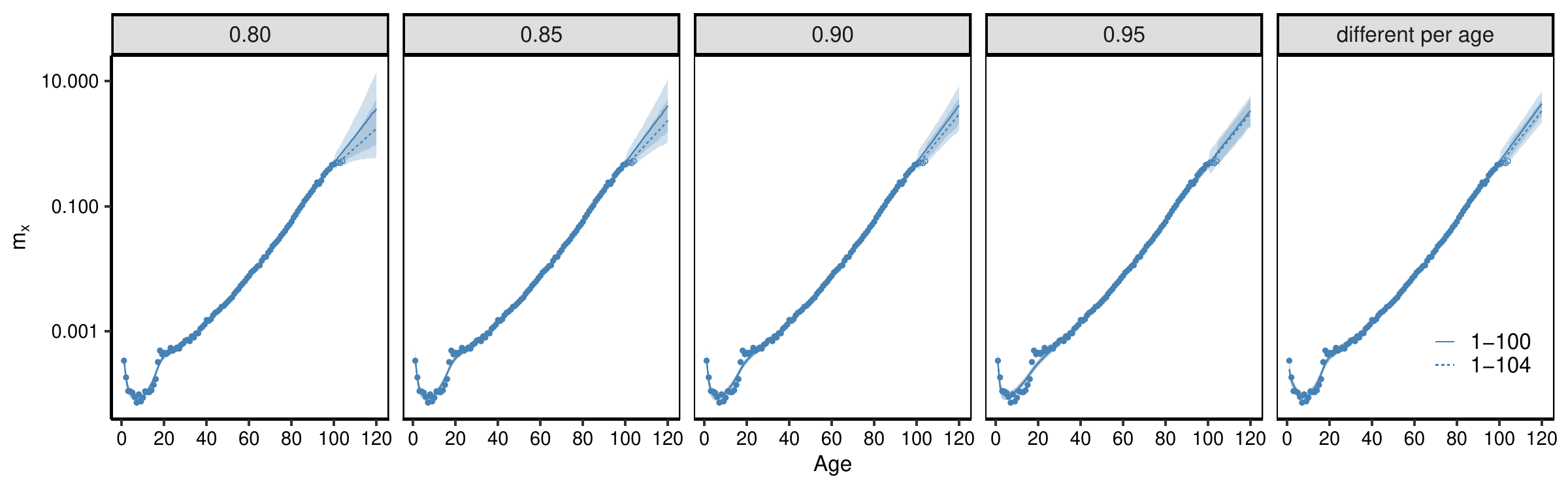}\\
    \end{tabular}
    \caption{Study 1 (E+W, univariate, {\bf male}, 2010-2012): the DLM fits extrapolated to the age of 120 with 95\% predictive credible interval and discount factor $\delta=(0.80, 0.85, 0.90, 0.95)$ and different discount factors per age ($\delta_{1,1:5}= 0.99$, $\delta_{2,6:35}=0.80$, $\delta_{3,36:85}=0.85$, $\delta_{4,86+}=0.99$). The solid line represents the mortality curve fit from ages 1-100 and the dashed line, from ages 1-104. First row: 2010, second row: 2011, third row: 2012.  }
    \label{fig:fig2}
\end{figure}    

Multivariate models allow for the capture of detailed interactions among multiple variables and their simultaneous influences on mortality rates, offering greater flexibility and better fit to observed data. This results in more precise and reliable predictions, reflecting a more realistic expectation of the convergence of the mortality rate between men and women, as age advances. In the following section, we consider jointly modelling both subpopulations.
\begin{figure}[H]
    \centering
    \begin{tabular}{c}
    \includegraphics[width=1\textwidth]{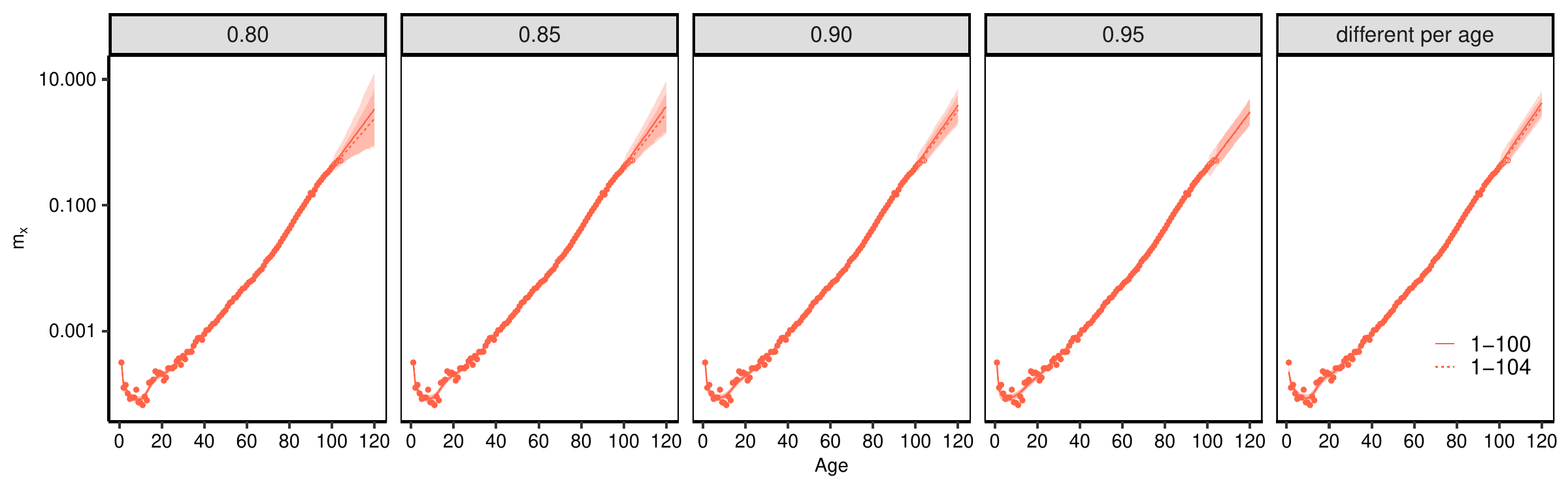}\\
       \includegraphics[width=1\textwidth]{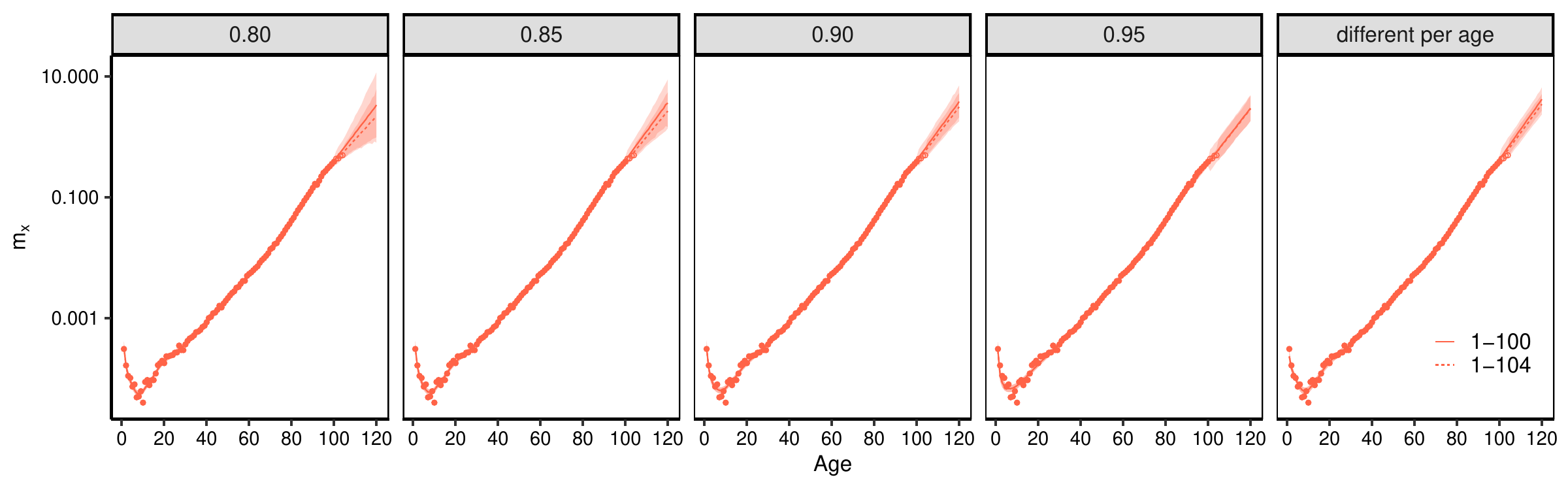}\\
          \includegraphics[width=1\textwidth]{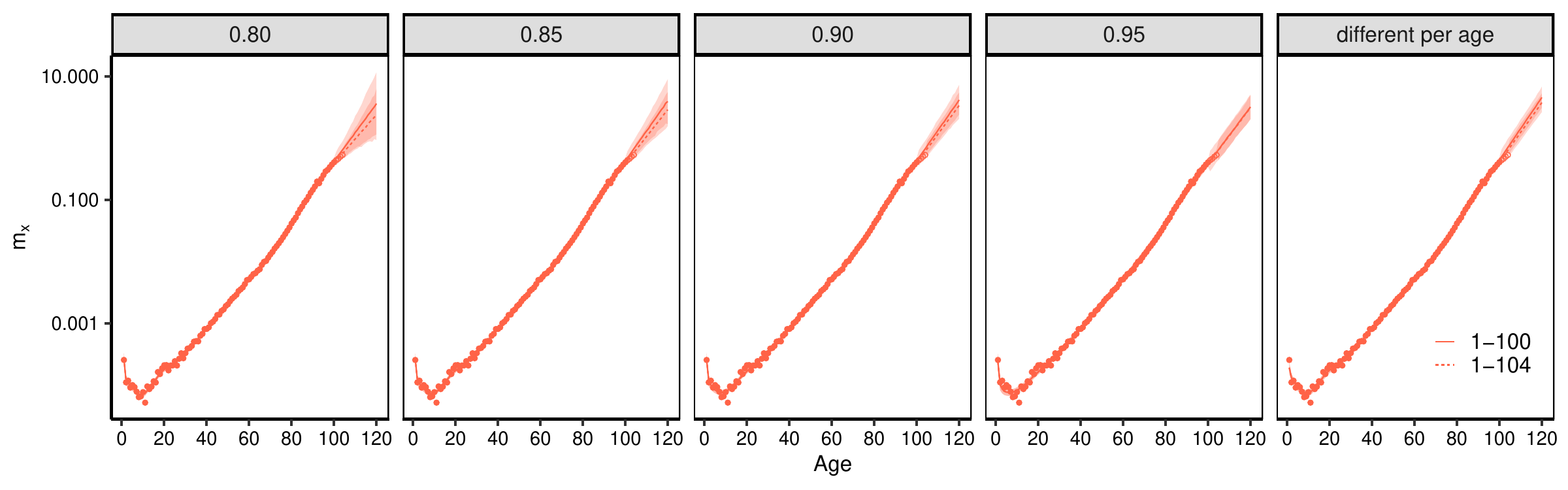}\\
    \end{tabular}
    \caption{Study 1 (E+W, univariate, {\bf female}, 2010-2012): the DLM fits extrapolated to the age of 120 with 95\% predictive credible interval and discount factor $\delta=(0.80, 0.85, 0.90, 0.95)$ and different discount factors per age ($\delta_{1,1:5}= 0.99$, $\delta_{2,6:35}=0.80$, $\delta_{3,36:85}=0.85$, $\delta_{4,86+}=0.99$). The solid line represents the mortality curve fit from ages 1-100 and the dashed line, from ages 1-104. First row: 2010, second row: 2011, third row: 2012. }
    \label{fig:fig3}
\end{figure}    

\begin{figure}[H]
    \begin{center}
    \begin{tabular}{c}
    \includegraphics[width=1\textwidth]{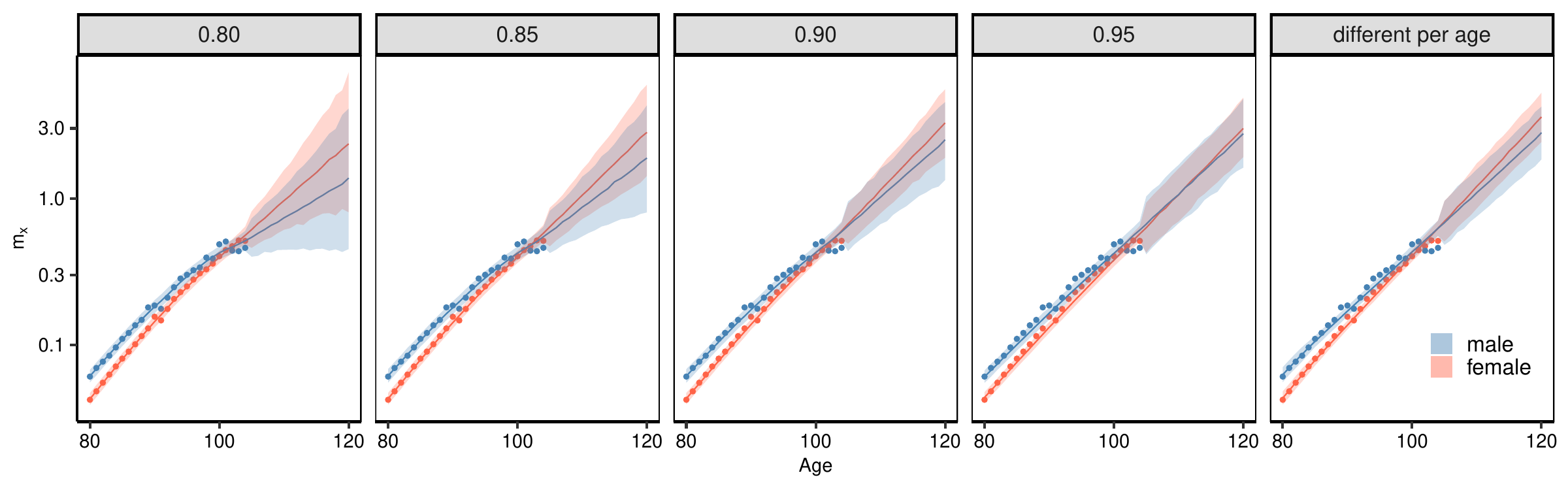}\\
       \includegraphics[width=1\textwidth]{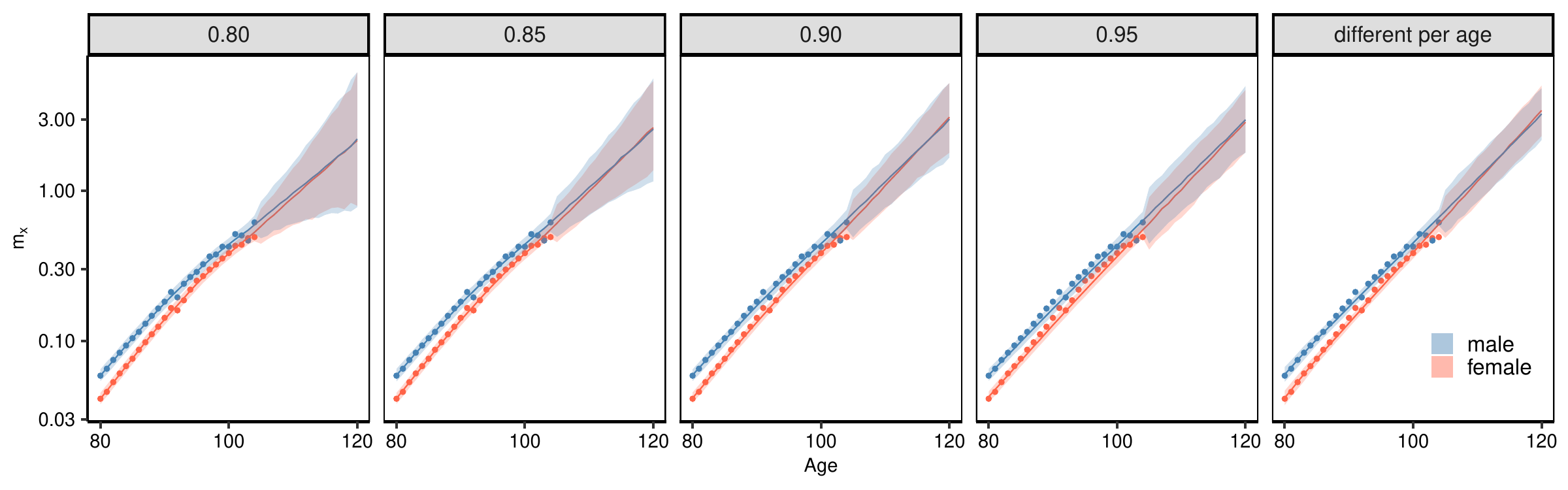}\\
              \includegraphics[width=1\textwidth]{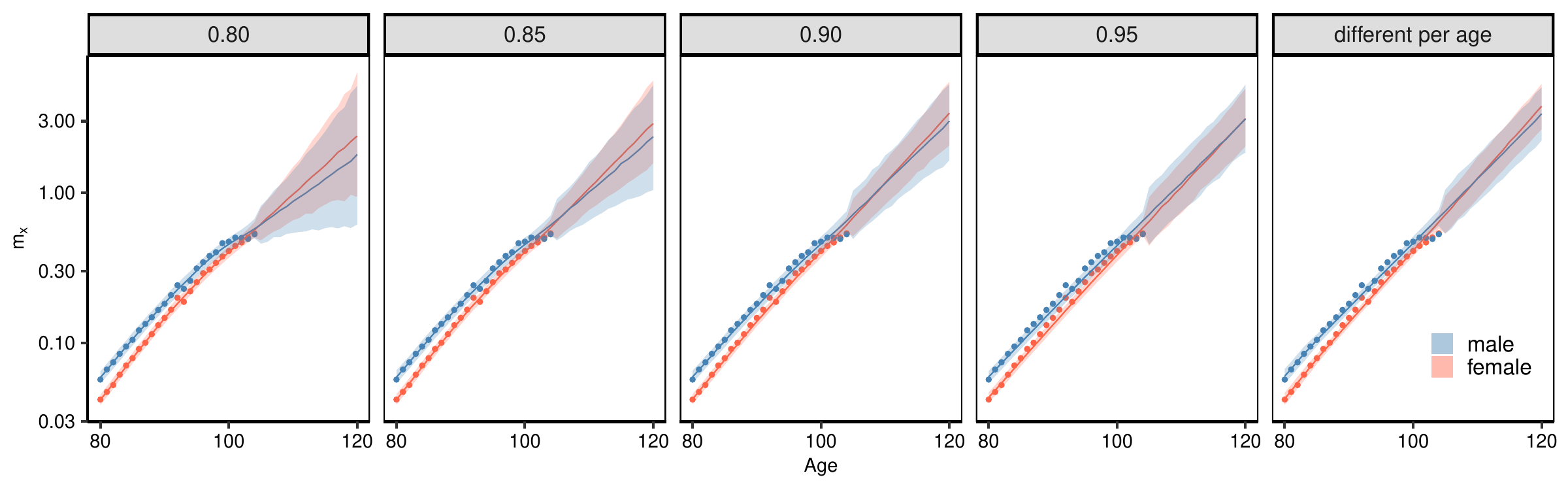}\\
    \end{tabular}
    \end{center}
     \caption{Study 1, scenario (a) (E+W,  {\bf univariate}, 80-104): the DLM fits extrapolated with 95\% predictive interval to the age of 120 with discount factor $\delta=(0.80, 0.85, 0.90, 0.95)$ and different discount factors per age ($\delta_{1,1:5}= 0.99$, $\delta_{2,6:35}=0.80$, $\delta_{3,36:85}=0.85$, $\delta_{4,86+}=0.99$), for the male and the female population for years 2010 (first row), 2011 (second row) and 2012 (third row). }
    \label{fig:fig4}
\end{figure}    

\begin{figure}[H]
    \begin{center}
    \begin{tabular}{c}
    \includegraphics[width=1\textwidth]{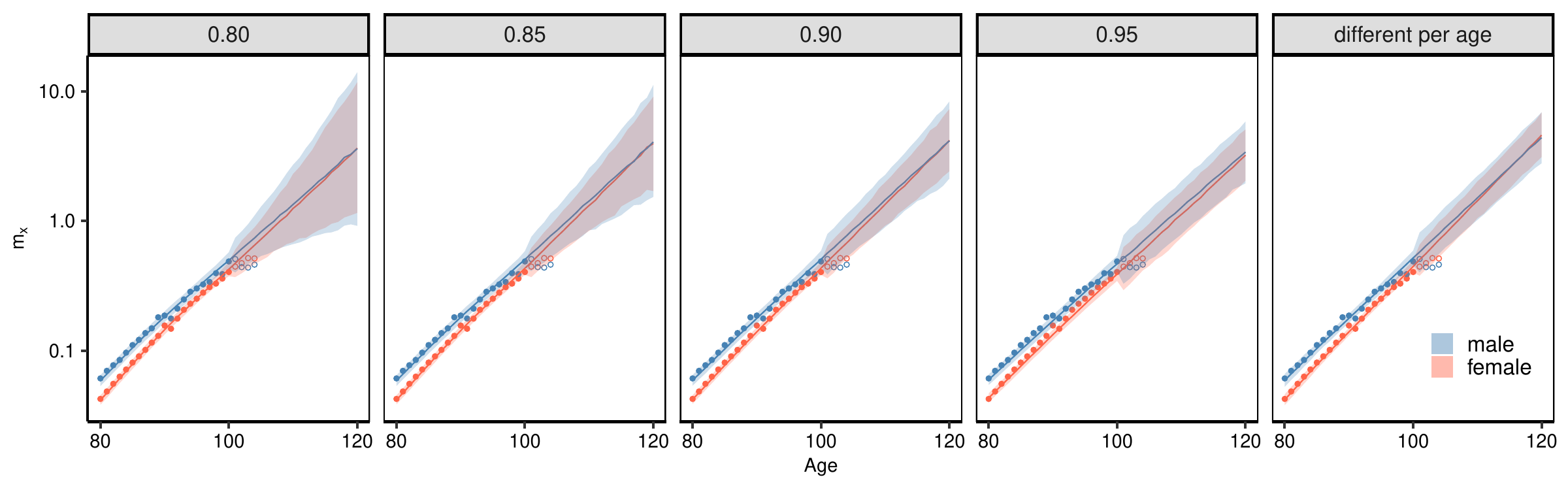}\\
       \includegraphics[width=1\textwidth]{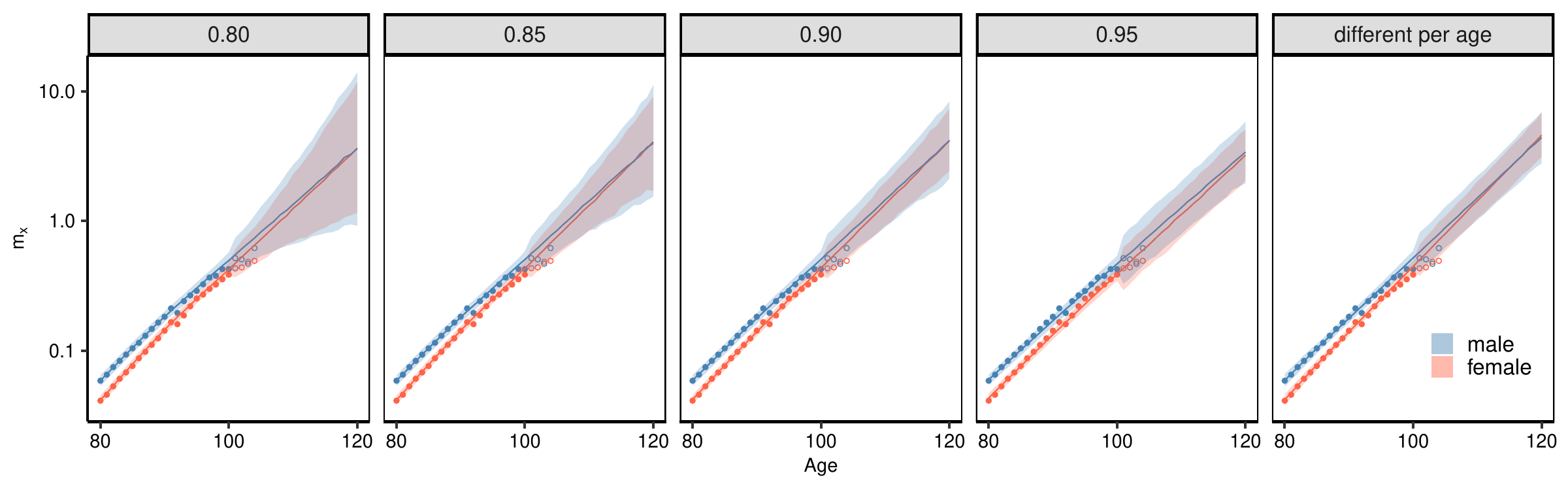}\\
              \includegraphics[width=1\textwidth]{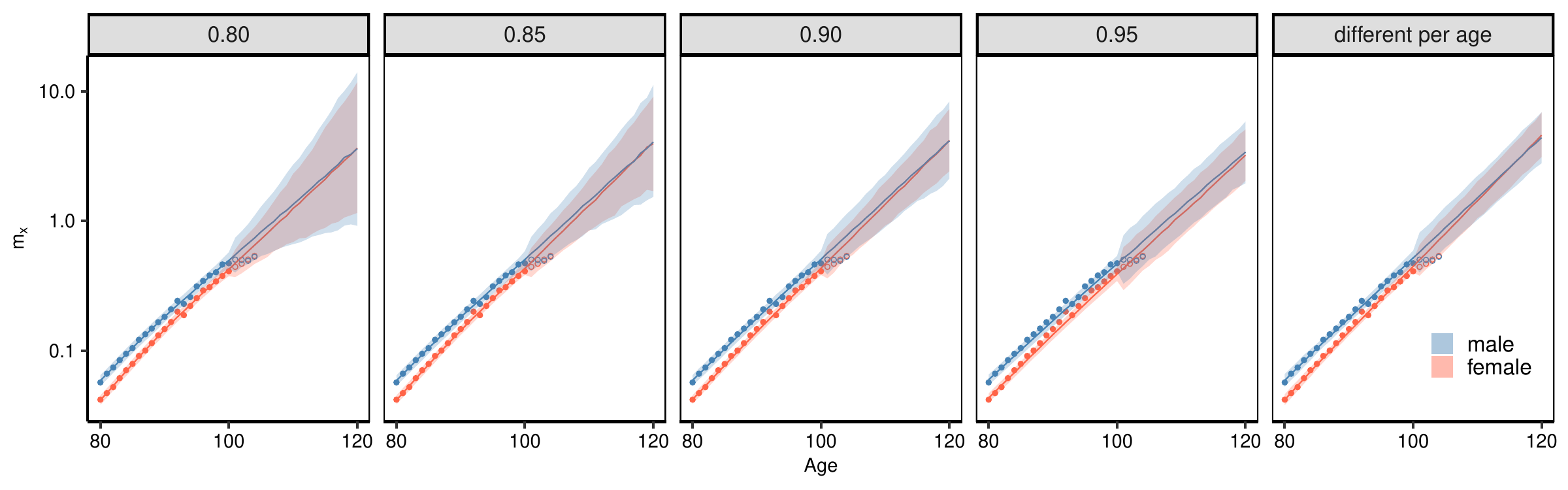}\\
    \end{tabular}
    \end{center}
   \caption{ Study 1, scenario (b) (E+W,  {\bf univariate}, 80-100): the DLM fits extrapolated with 95\% predictive interval to the age of 120 with discount factor $\delta=(0.80, 0.85, 0.90, 0.95)$ and different discount factor per age ($\delta_{1,1:5}= 0.99$, $\delta_{2,6:35}=0.80$, $\delta_{3,36:85}=0.85$, $\delta_{4,86+}=0.99$), for the male and the female population for years 2010 (first row), 2011 (second row) and 2012 (third row).}
    \label{fig:fig5}
\end{figure}    

\subsection{Study 2: Joint modelling - convergence and prevention of cross-over }\label{sec4.2}

We consider the same dataset analysed in Study 1. Here, our goal is to illustrate the benefits of jointly modelling male and female mortality rates. As demonstrated in the previous study by \cite{forster22}, using adaptive P-splines models, information can be exchanged between sexes, considerably improving model fit for males, particularly at advanced ages, where there tends to be a larger amount of female data available. Furthermore, the issue of overlapping male and female mortality rates can be mitigated. In our proposed model given by equations (\ref{eq.obs1}), (\ref{eq.evol.mu1}) and (\ref{eq.evol.beta1}), this is achieved by introducing a constraint in the modelling process and incorporating the use of discount factors.

In this study, we consider the age range from 1 to 104 for the modelling process. As presented in Study 1, the adoption of age-varying factors yields satisfactory predictions with robust uncertainty quantification, without constraining the age range.  Additionally, it is expected that the joint modelling of subpopulations will better handle the estimation at older ages mortality compared to the marginal models (separate modelling for males and females). As stated in the previous study, males and females mortality curves have the same age-varying smoothness, although the proposed model is capable of accommodating different $\delta^{(j)}_x$ per subpopulation. 

Figure \ref{fig:fig6} presents the results of the joint model fit considering all ages (1-104) and the zoomed-in versions (1-30) and (80-104 +, projected up to age 120) for the 2010-2012 male and female populations. As evidenced in the analysis, the flexibility achieved by using age-varying factors $\delta_x$ along with joint modelling results in a better fit over the years compared to using unique discount factors (similar results were seen in Study 1 and are therefore omitted here). {The convergence between the subpopulations occurs smoothly, with the terminal age for convergence of mortality curves set to 105 years old for 2010 (since the male and female curves had already crossed by this age) and 115 years for 2011 and 2012}. Additionally, the joint modelling of mortality curves reduces the uncertainty associated with the estimates in the extrapolation process.

\begin{figure}[H]
    \centering
    \begin{tabular}{r}
\includegraphics[width=11cm, height=3.9cm]{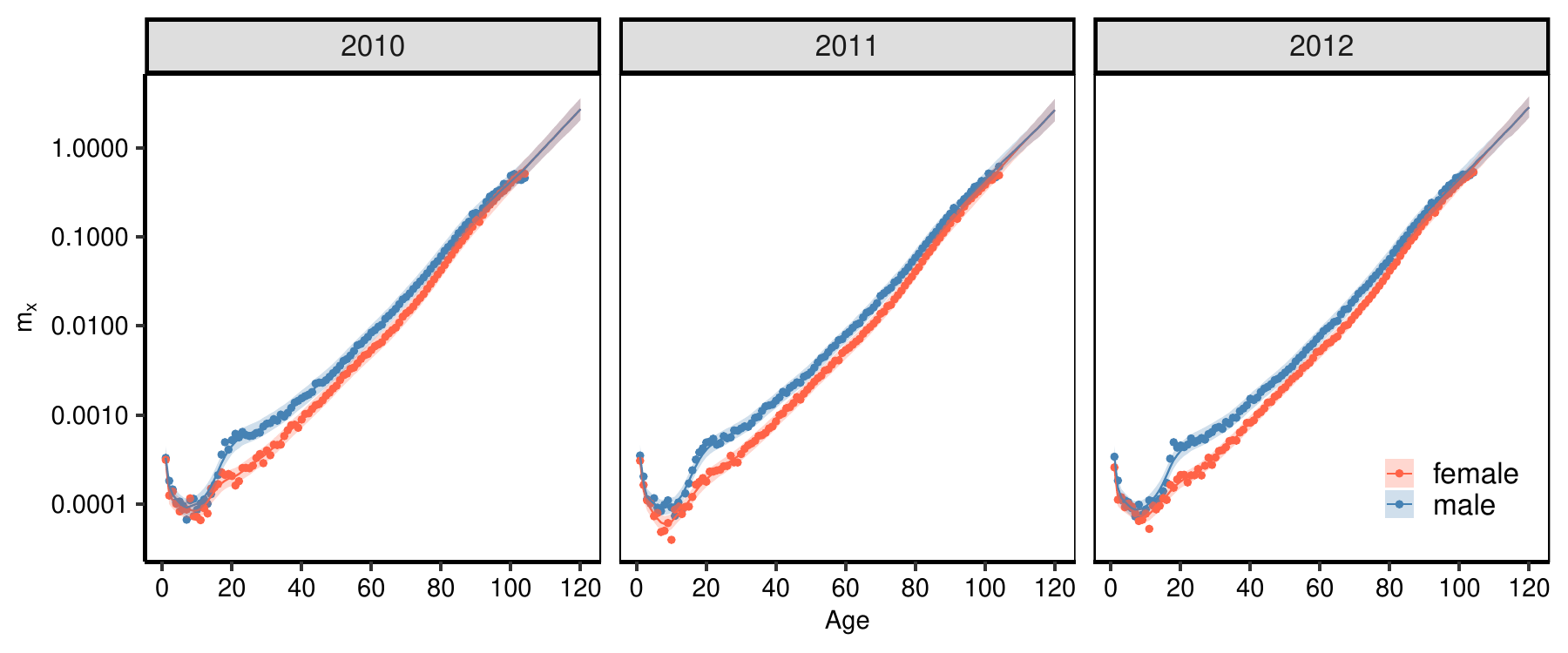}\\
\includegraphics[width=10.8cm, height=4cm]{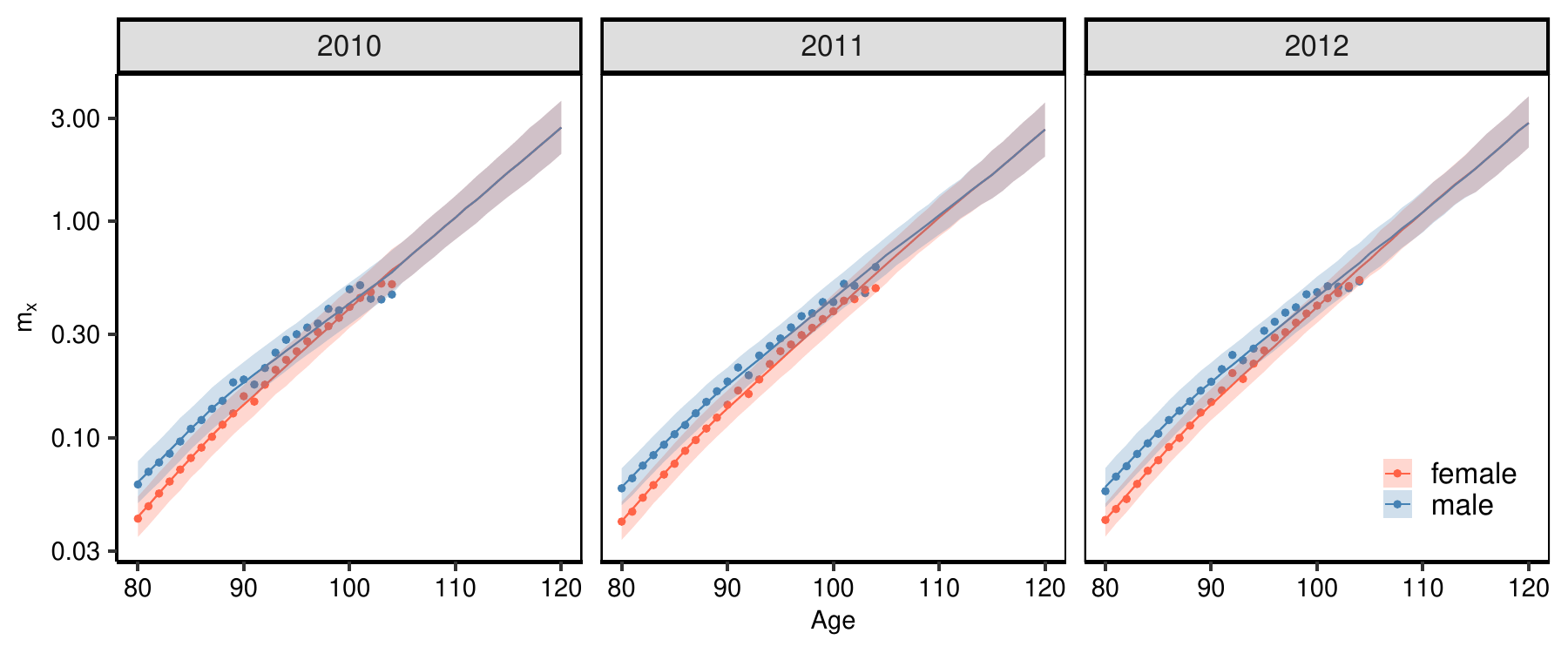}\\
          \includegraphics[width=11.1cm, height=4cm]{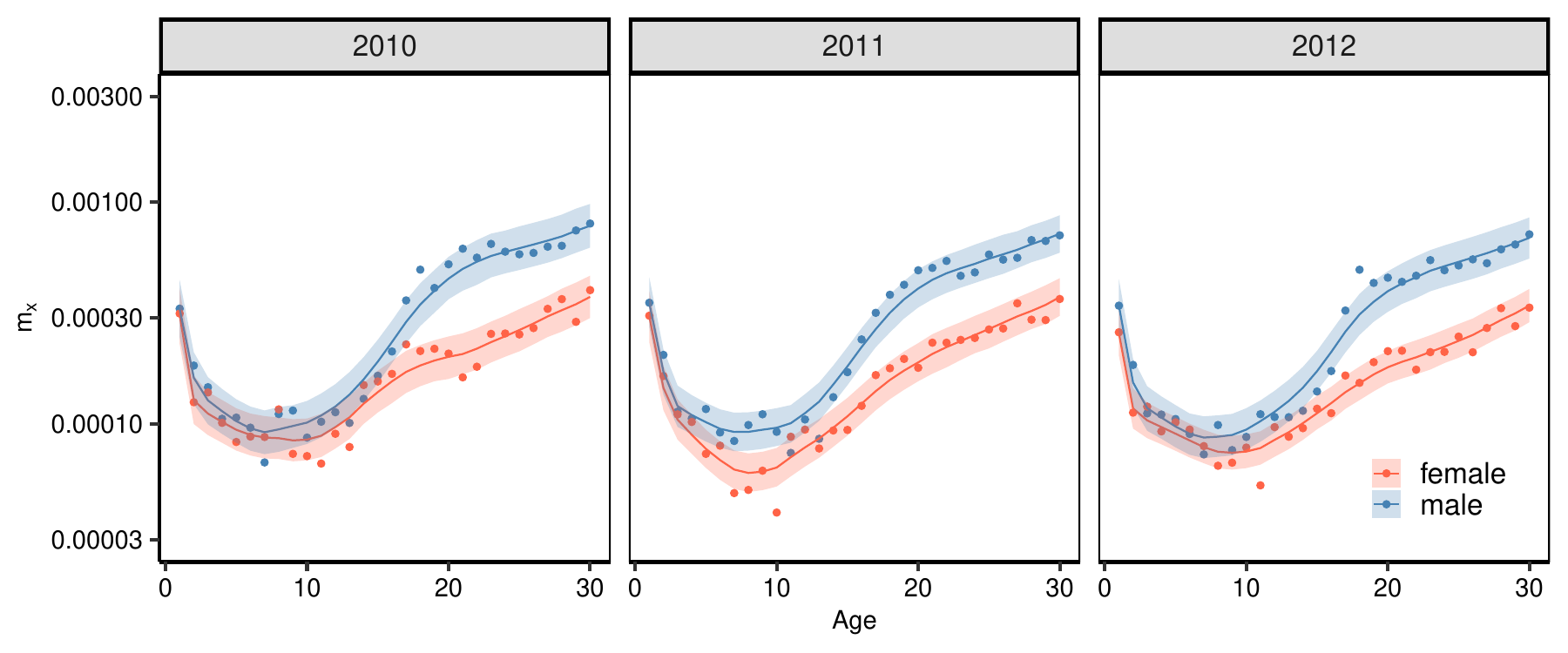}\\
    \end{tabular}
\caption{Study 2 (E+W,  {\bf bivariate}, 1-104): the DLM fits extrapolated with 95\% predictive interval up to  the age of 120, applying age-varying $\delta_x$ and  considering convergence between male and female populations over 2010-2012.}    \label{fig:fig6}
\end{figure}    

{Although not shown here, it is worth noting that the curve estimated using the model with a common factor, as described in Section \ref{sec3}, exhibits a similar behaviour to the traditional bivariate model. However, when extrapolating, the associated uncertainty is higher, as all available data are used in the graduation process. In this setting, the common factor tends to resemble the more robust population — in this study, the male population — leading to extrapolated behaviour across all ages that may differ from the bivariate model without a common term.}

\subsubsection{Sensitivity analysis on prior specification and age-varying smoothness for cross-over prevention}

{Although our study does not show crossing between the fitted male and female mortality curves at infant and young ages for the years 2010 to 2012, there is no guarantee that the estimated male mortality rates will consistently exceed the female rates (see Figure \ref{fig:fig6}, third row). In Tang's study, the 2012 mortality fit exhibits a slight crossover at infant ages between populations, prompting the addition of a penalty on the difference between male and female P-spline coefficients to prevent this issue. To address this in our model specification, we perform a sensitivity analysis on the prior settings for the state parameters ($\alpha_x$ and $\beta_x$) and the age-varying smoothness parameter ($\delta_x$), aiming to improve the estimation of initial-age mortality patterns.} 

{To illustrate, we focus on the year 2012, during which a cross-over between the populations was observed in Tang’s study. We consider both fixed discount factors across ages and age-varying factors, as already examined in Study 1. Additionally, we propose three values for the hyperparameter $\bm C_{x_0}$ to control the amount of prior information, ranging from a more informative to a more vague specification. In this context, an informative prior can help stabilise the model and may reduce the risk of cross-over. On the other hand, a vague prior offers flexibility, although it may lead to less reliable estimates, especially when the data are noisy.}

{Figure \ref{fig:bivar_crossover} 
displays the results of combining various prior specifications with discount factors for the multivariate models discussed in Section \ref{sec3}. Comparing weakly informative priors (e.g., $\bm C_{x_0} = \text{diag}(100)$) with more informative ones (e.g., $\bm C_{x_0} = \text{diag}(1)$) reveals a reduction in posterior uncertainty in the latter case, particularly at younger ages. This reduction can affect the distance between the fitted curves, either bringing them closer or pushing them apart, depending on the strength of the prior information and the chosen discount factor.}{ In this context, it may be of particular interest to perform a sensitivity analysis to assess whether more informative priors, in combination with age-varying discount factors, improve the model fit by reducing uncertainty and helping to prevent cross-overs. Although not displayed, the joint model with a latent common term yielded nearly identical results.

\begin{figure}[H]
    \centering
    \includegraphics[width=0.8\textwidth]{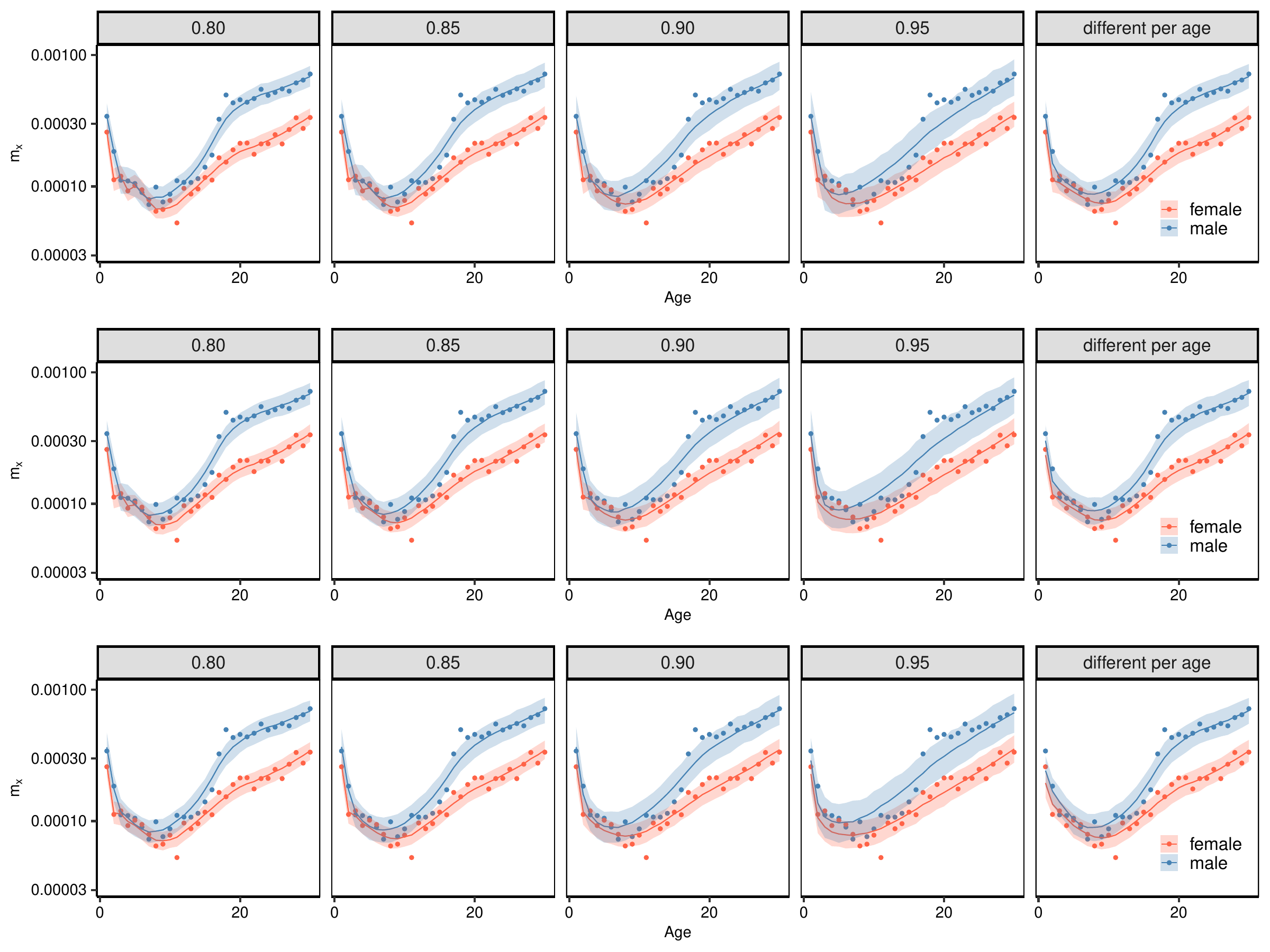}\\
\caption{Study 2 (E+W,  {\bf bivariate}, 1-104): the DLM fits with 95\% predictive interval, applying fixed  $\delta=(0.80, 0.85, 0.90, 0.95)$ and age-varying discount factors  $\delta_x$,  varying prior variance $C_{x_0} = \text{diag}(100)$ (first row), $C_{x_0} = \text{diag}(10)$ (second row) and $C_{x_0} = \text{diag}(1)$ (third row), for the male and female populations, year 2012.}
    \label{fig:bivar_crossover}
\end{figure}


\subsection{Study 3: Joint modelling for missing data}\label{study3}

The results presented in Sections \ref{sec4.1} and \ref{sec4.2} offer a valuable understanding of mortality behaviour and assess the performance and accuracy of different modelling approaches. In this section, we revisit the dataset from \cite{forster22} to create scenarios with incomplete information for the female population. Focusing on the year 2010 (ages 1-104), we construct various scenarios by selectively removing mortality data, particularly from younger groups, 
as outlined in Table \ref{st3:tab1}. These scenarios are designed to test the robustness of the proposed model when faced with incomplete data across different age ranges.

 To address the lack of data in different age groups, a feasible approach is to take into account information from the population that provides robust data (in our scenario, the male one) to borrow information for the population with missing data (female). By doing so, we can improve the estimation of parameters, especially in age groups for which data are scarce or non-existent. This approach can be particularly beneficial when the mortality data for one of the populations in the study is more complete and detailed. Furthermore, this method can be applied in any context where there is a need to fill information in a life table.
\begin{table}[h!]
\centering
\caption{Study 3: scenarios for missing mortality data.}
\label{st3:tab1}
\begin{tabular}{|c|c|c|}
\hline
\textbf{scenario} & \textbf{age groups} & \textbf{$\%$ of missing data} \\
\hline
a & 4-8 (Female) & $\approx 5\%$ \\
b &  4-10, 15-17 (Female) & $\approx 10\%$ \\
c & 3-16 (Female) & $\approx 15\%$ \\
d & 1-25 (Female) & $\approx 25\%$ \\
e & 1-16, 23-41 (Female) & $\approx 33\%$ \\
f & 1-45 (Female) & $\approx 43\%$ \\
\hline
\end{tabular}
\end{table}

Two competing models are considered here: the usual joint model, and the joint model with a common term, both proposed in Section \ref{sec3}. 
The same age-varying discount factors $\delta_x$ are applied for both populations. Figure \ref{fig:fig10} exhibits the resulting fits from both models considering the complete data. As shown, the models perform well, providing results that are consistent, as expected.

\begin{figure}[H]
\centering
\begin{tabular}{c}
\includegraphics[width=0.5\textwidth]{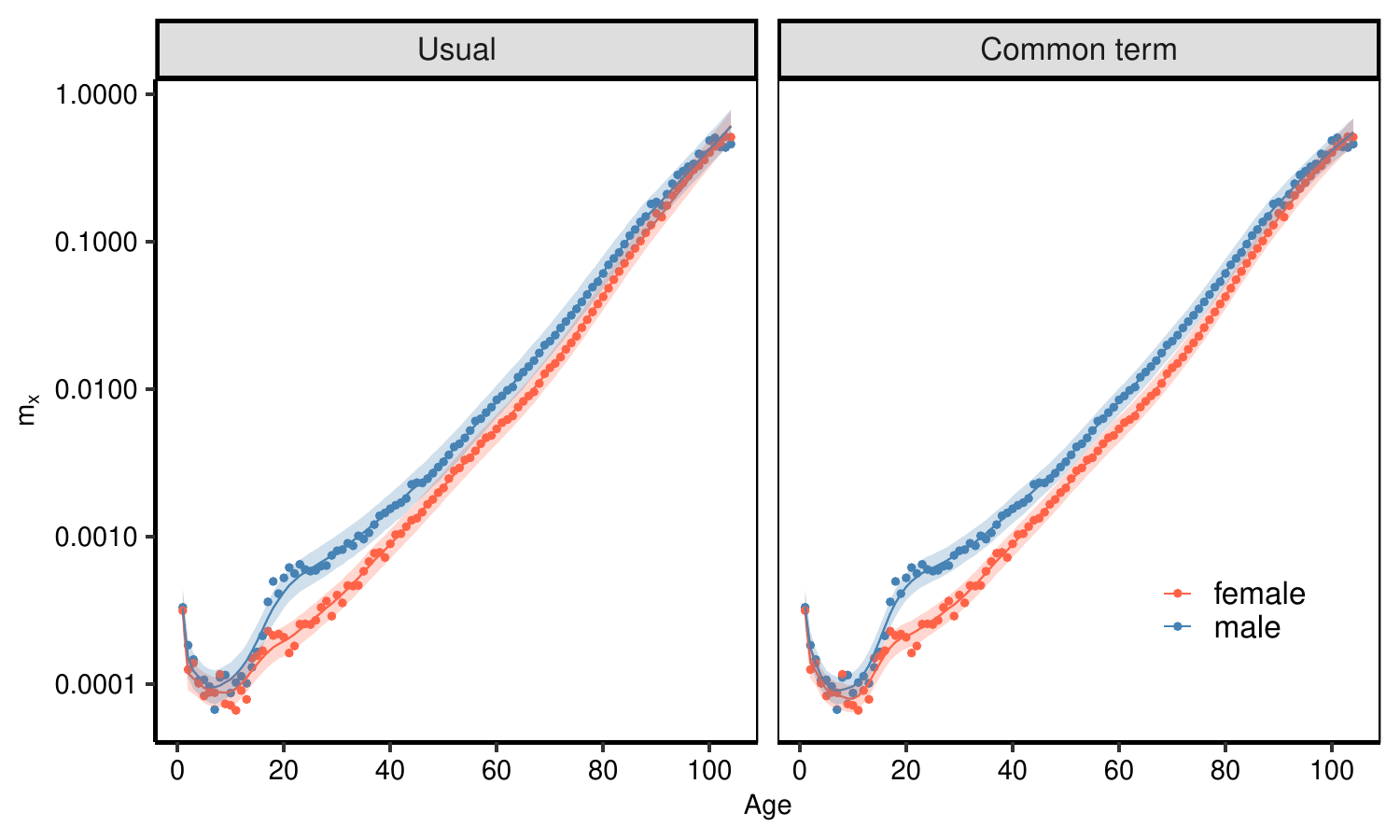}\\
\end{tabular}
\caption{Study 3 (E+W, {\bf no missing data}, 1-104): the DLM fits (with discount factors $\delta_x$)  and 95\% predictive interval. Solid lines represent the fit considering the complete dataset for the joint model (left panel) and the joint model with common term (right panel).}
\label{fig:fig10}
\end{figure}

The performance of the  proposed models in estimating mortality curves in the presence of missing data was evaluated. For this, we estimated the log-mortality curves using the complete dataset (represented by black dashed lines in the following figures) and compared them with the fitting scenarios involving missing data (solid lines) considered in Table \ref{st3:tab1}. The red and blue shaded areas show 95\% predictive intervals for female and male groups, respectively, and the empty dots indicate missing data. The uncertainty around the log-mortality estimates decreases with the increased amount of log-mortality information.

Zoomed panels in Figure \ref{fig:fig11} display scenarios from (a) to (f) for both models. The aim is to vary the percentage of missing data in the female dataset and to understand how data gaps during childhood and young ages affect the overall mortality curve, as well as how pooling can help mitigate these effects. As can be seen, all joint models are able to capture the behaviour of the true mortality curve when less than 10\% of the data are missing. 
In scenario (c)  with approximately 15\% missing data in infant ages, the usual joint model does not provide good performance, resulting in cross-over curves in the region where data are unavailable. However, the introduction of the common term could be an alternative by mimicking the male mortality behaviour in the region where there are gaps in the female group. 
\begin{figure}[H]
\centering
\begin{tabular}{cc}
\includegraphics[width=0.5\textwidth]{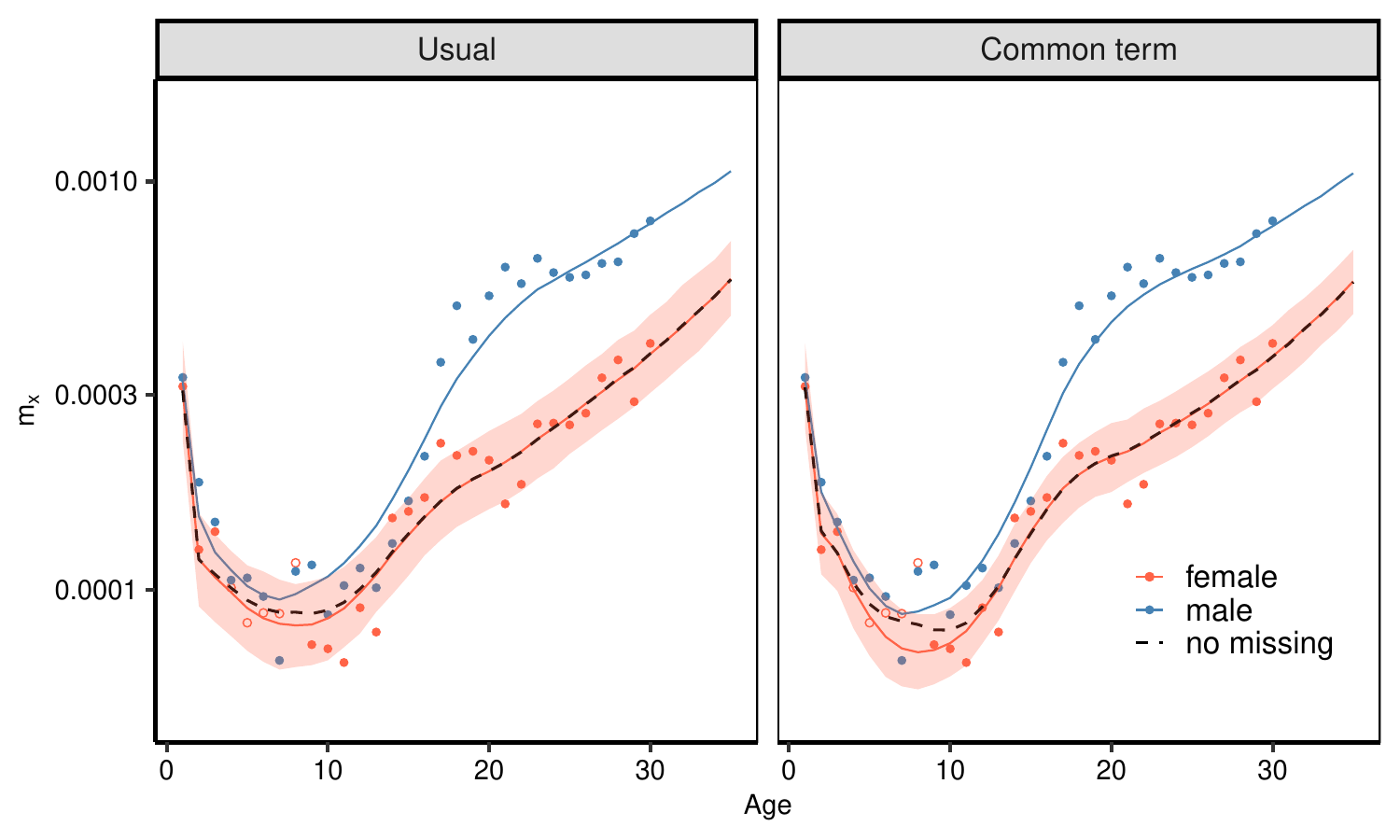} &  \includegraphics[width=0.5\textwidth]{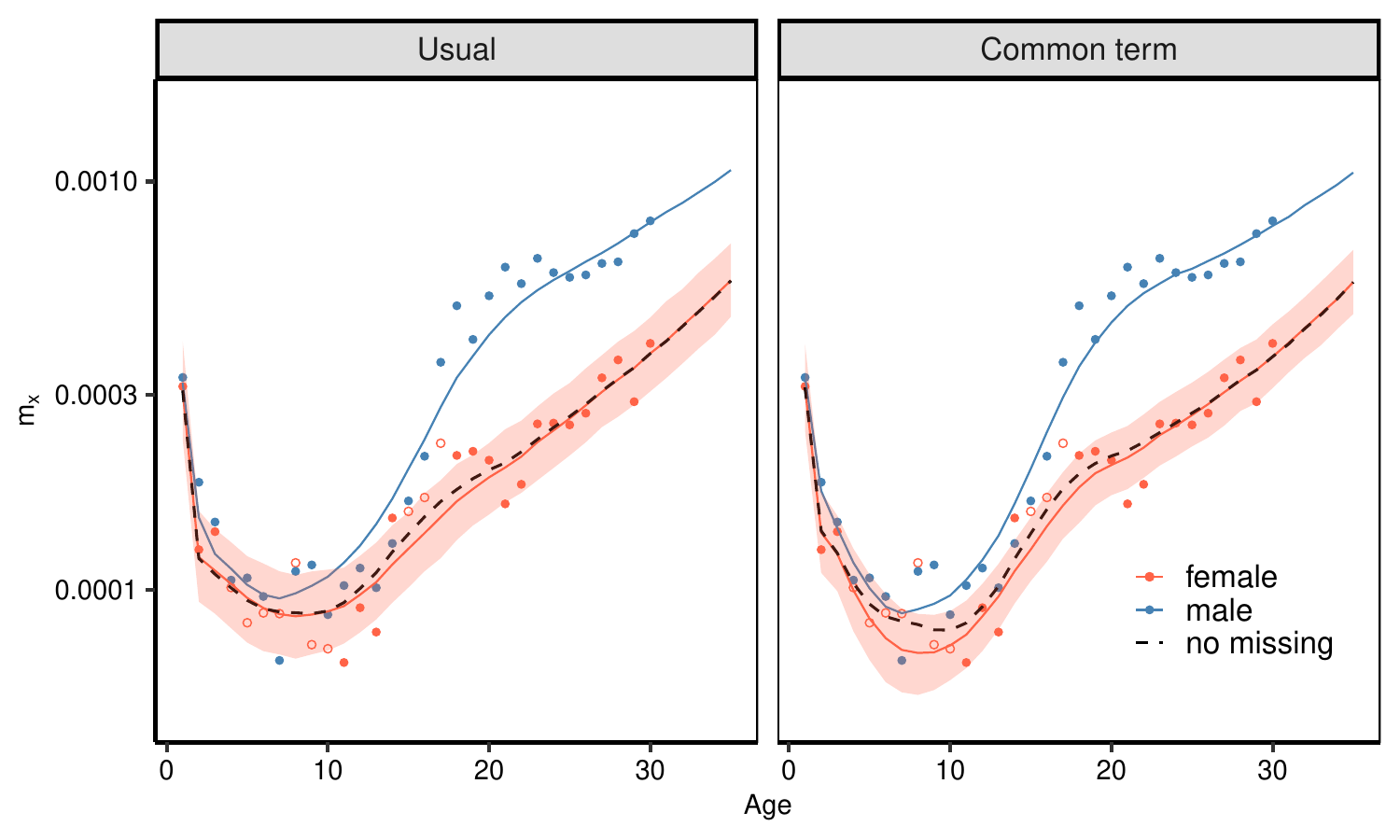}\\
scenario (a) & scenario (b) \\
\includegraphics[width=0.5 \textwidth]{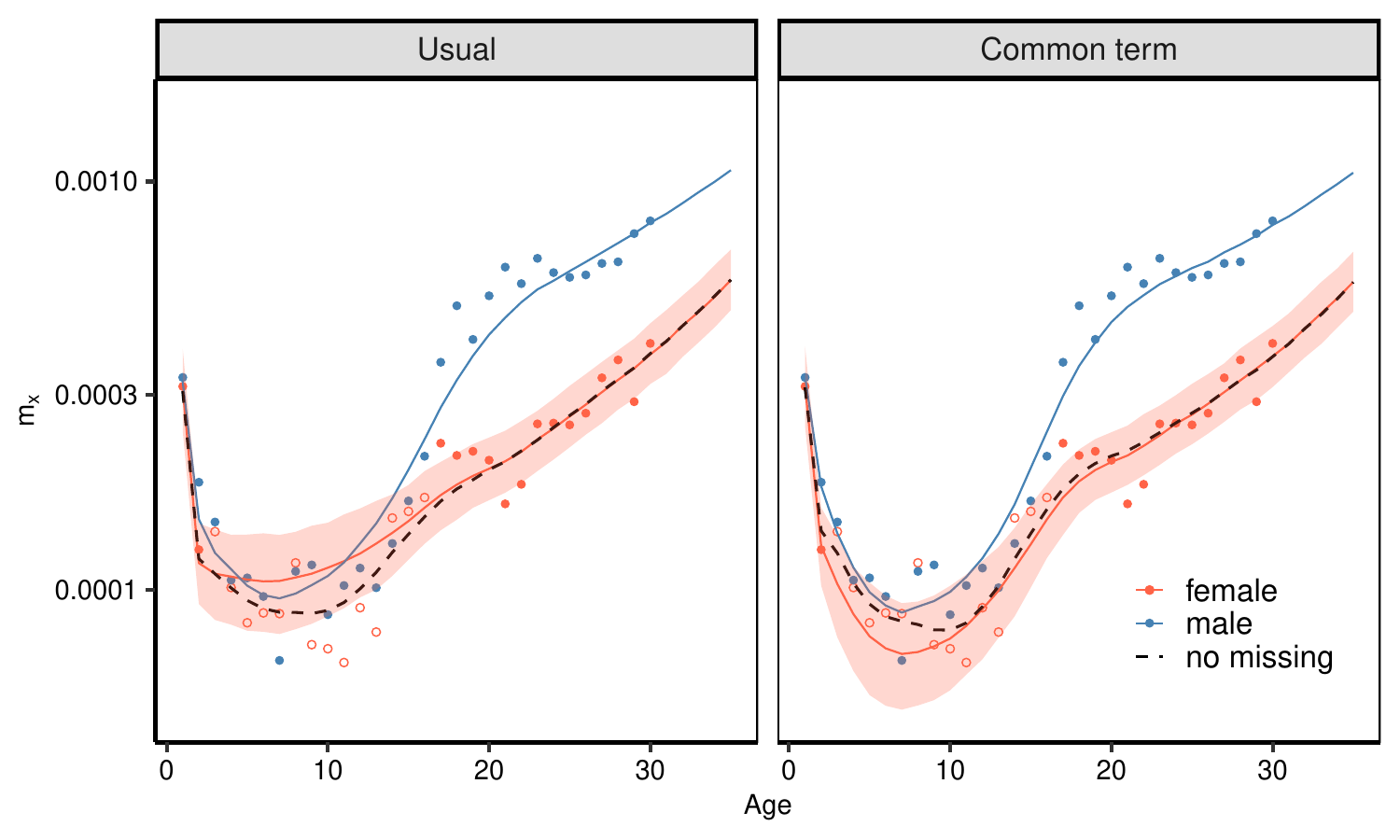} & \includegraphics[width=0.5\textwidth]{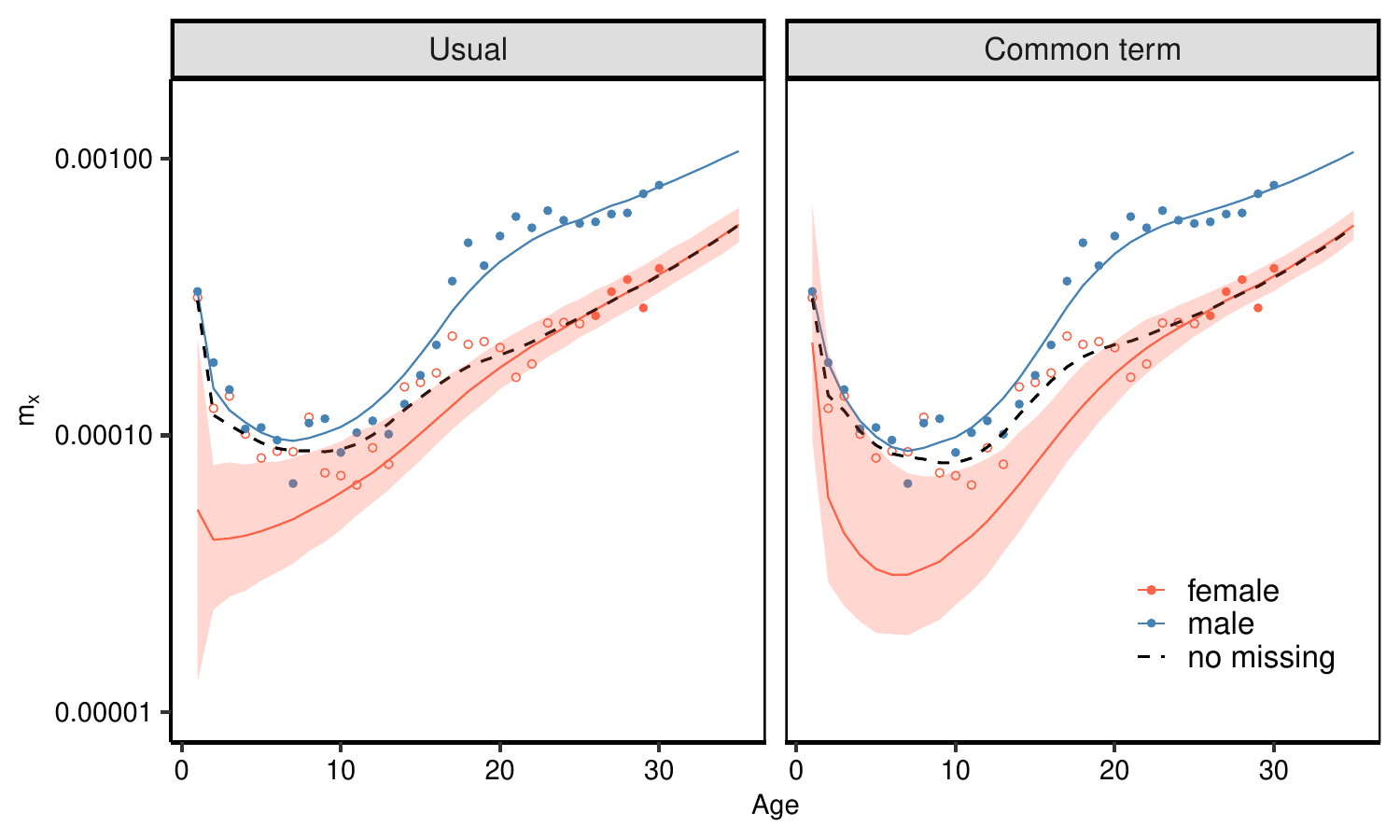} \\
scenario (c) & scenario (d) \\
\includegraphics[width=0.5\textwidth]{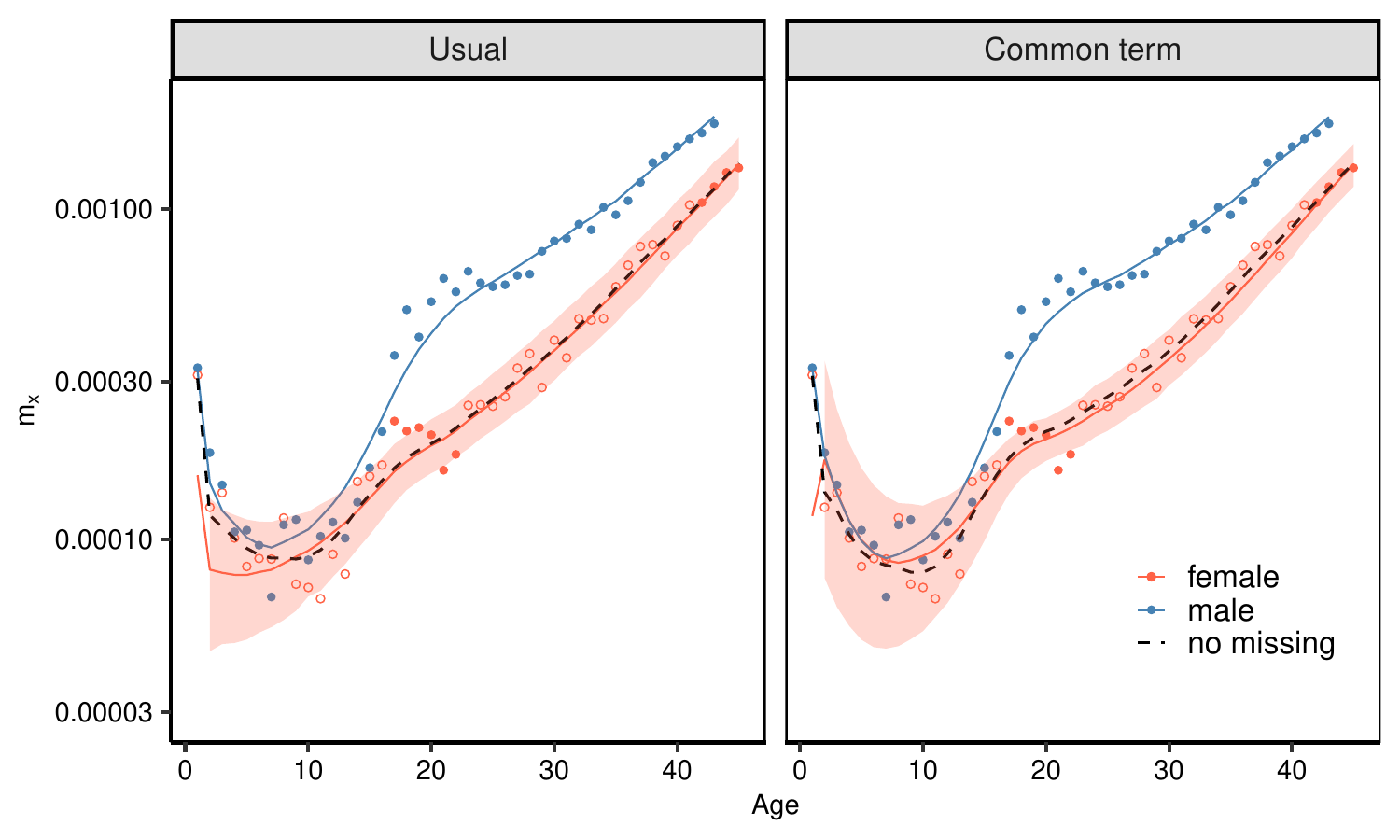} & 
\includegraphics[width=0.5\textwidth]{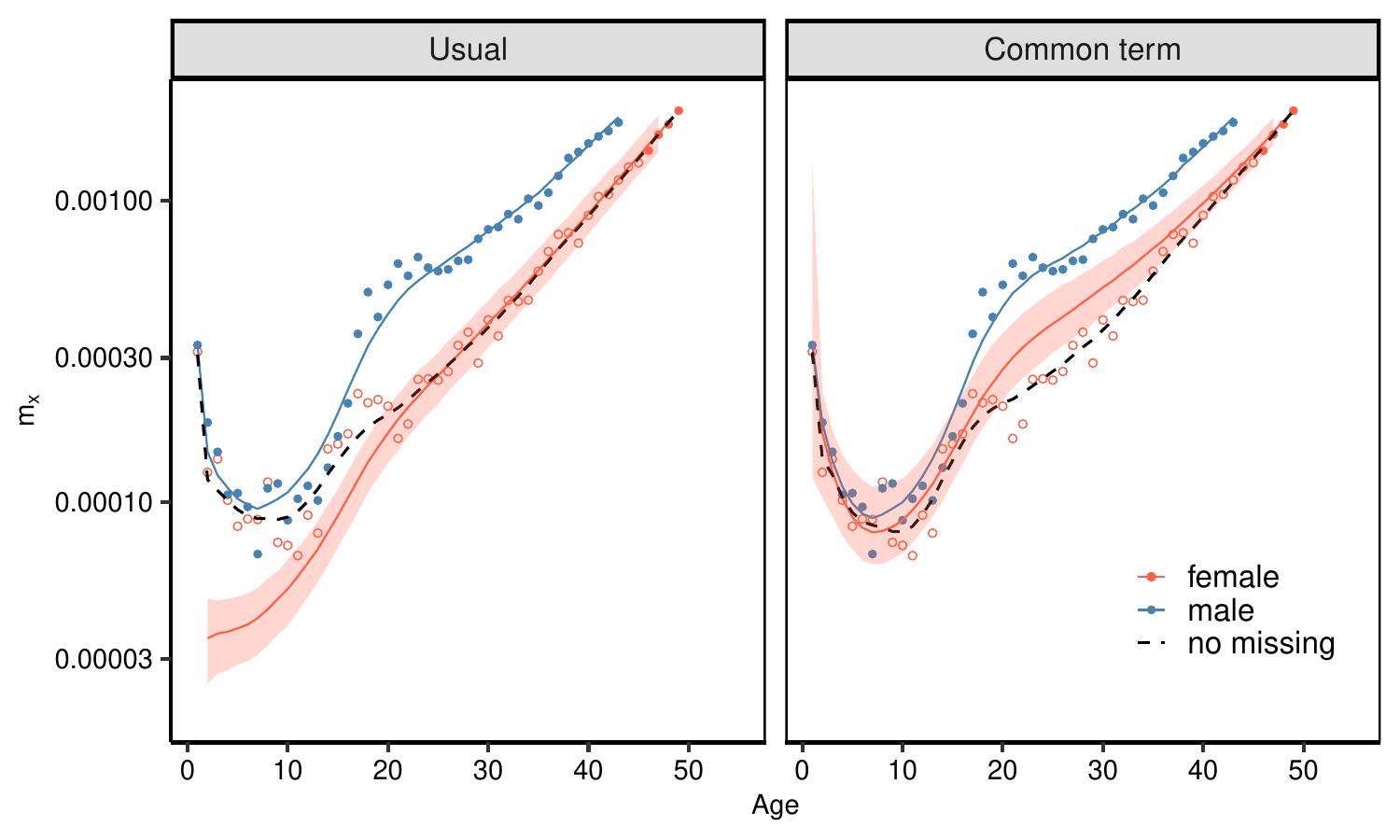}\\
scenario (e) & scenario (f) \\
\end{tabular}
\caption{Study 3 (E+W, {\bf missing data} missing data in the female population, scenarios from (a) to (f)): the DLM fits with discount factors $\delta_x$ and 95\% predictive interval. Empty dots indicate missing data. Black dashed lines represent the fit considering the complete dataset (no missing).}
\label{fig:fig11}
\end{figure}

In the following scenarios (d) to (f), we consider higher levels of missing data starting from the earliest age, which could be more challenging for the models to capture a plausible behaviour of the mortality curve. The lack of information, particularly in younger adult ages (accident hump), means that the models lack guidance on the true level of the curve. Notice that in scenario (e) where this information is available -- between 17-22 years old -- the performance is better compared to scenarios (d) and (f). Once again, the model with the common term appears to be quite effective in filling in data in this age range. Moreover, models with a common term can correctly increase uncertainty in the area with too high levels of missing data. Table \ref{st3:tab2} shows the measures for the comparison of models using MSPE, MAPE, and WCI (see Appendix \ref{apB} for more detailed information on the measures). The measures highlighted in bold indicate the best performance. As expected, models without a common term present the worst predictive performance in the presence of higher levels of missing data.

\begin{table}[ht]
\centering
\caption{Study 3: Model comparison based on Mean Square Prediction Error (MSPE), Mean Absolute Prediction Error (MAPE) and Width of Credibility Interval (WCI) for scenarios (a)-(f) a varying percentage of missing data for both competing models.}
\label{st3:tab2}
\begin{tabular}{|l|c|c|ccc|}
\hline \hline
\multirow{2}{*}{{\bf competing models}} &  \multirow{2}{*}{\bf{scenario}} &\multirow{2}{*}{\bf{\% missing}} & \multicolumn{3}{c|}{{\bf measures}} \\ 
                       &                   &          & mspe   & mape    & wci    \\ \hline
Bivariate usual       & \multirow{2}{*}{a}& \multirow{2}{*}{$\approx 5\%$}  & 0.00097 & 0.00853 & 0.02524 \\ 
Bivariate w/ CT      &  &                                & {\bf0.00062} & {\bf0.00772} & {\bf 0.02056} \\ 
\hline\hline
Bivariate usual        &\multirow{2}{*}{b}& \multirow{2}{*}{$\approx 10\%$} & 0.00097 & 0.00850 & 0.02414 \\ 
Bivariate w/ CT        &&                                & {\bf0.00063} & {\bf0.00777} & {\bf0.02021} \\ 
\hline\hline
Bivariate usual        &\multirow{2}{*}{c}& \multirow{2}{*}{$\approx 15\%$} & 0.00099 & 0.00855 & 0.02017 \\ 
Bivariate w/ CT        &&                                & {\bf0.00064} & {\bf0.00779} & {\bf0.01901} \\ 
\hline\hline
Bivariate usual        &\multirow{2}{*}{d}& \multirow{2}{*}{$\approx 25\%$} &0.00097 & 0.00848 & 0.01710 \\ 
Bivariate w/ CT        &&                                & {\bf0.00066} & {\bf0.00784} & {\bf 0.01392} \\ 
\hline\hline
Bivariate usual        &\multirow{2}{*}{e}& \multirow{2}{*}{$\approx 33\%$} &0.00098 & 0.00851 & 0.02063 \\ 
Bivariate w/ CT        &&                                & {\bf0.00066} & {\bf0.00787} & {\bf 0.01729} \\ 
\hline\hline
Bivariate usual        &\multirow{2}{*}{f}& \multirow{2}{*}{$\approx 43\%$} &0.00098 & 0.00854 & 0.01591 \\ 
Bivariate w/ CT        &&                                & {\bf 0.00075} & {\bf 0.00811} & {\bf 0.01159} \\ 
\hline
\end{tabular}
\end{table}

\subsubsection{Analysis of Sex Differences in Mortality Patterns}

Although the previous analysis focused on addressing missing data using male mortality rates, expanding the study to include other populations would allow for a more comprehensive understanding of mortality dynamics. Mortality curves for females and males exhibit distinct patterns, notably in the so-called hump. Male mortality typically shows a sharper increase during young and middle adulthood, largely due to external causes. In contrast, female mortality follows a smoother path, gradually increasing over time, reflecting differences in biological and behavioural factors.  By incorporating populations that capture this behaviour in female mortality data, the model could more accurately capture these gender-specific trends, particularly the female mortality hump, enhancing its overall robustness and applicability across populations, especially in contexts where the percentage of missing data in this age range is higher.

In addition to the 2010 male mortality data, female mortality data from 2012 are incorporated to provide supplementary information for the estimation of the 2010 female mortality curve. In this context, we assume that any population exhibiting a similar behaviour to the population with missing data can be used to fill in the missing values. For instance, two scenarios are considered based on Table \ref{st3:tab1}: a realistic scenario (c) with 15\% missing data and an extreme scenario (f) with 43\% missing data. In the following results, we present only the usual and common term models for comparison. To allow flexibility of mortality curve estimates, we apply a discount factor of 0.9995 for ages 1 to 5 keeping the same smoothness for the other ages as in the previous studies.

The zoomed-in panels in Figure \ref{fig:fig13} illustrate scenarios (c) and (f), considering the pooling with three populations. In scenario (c), the information borrowed across all 2010 female mortality curves enables the estimation of the missing data mortality curve closely resembling the true fit (assuming no missing data – see dashed line). In scenario (f), the common term model demonstrates behaviour more consistent with female patterns at infant and young ages compared to what was observed in Figure \ref{fig:fig11} (panel 6, column 2), where the information was derived only from the male population in 2010. Once again, the usual model fails to address the gaps in the missing mortality data.
\begin{figure}[H]
\centering
\begin{tabular}{c c}
\includegraphics[width=0.5\textwidth]{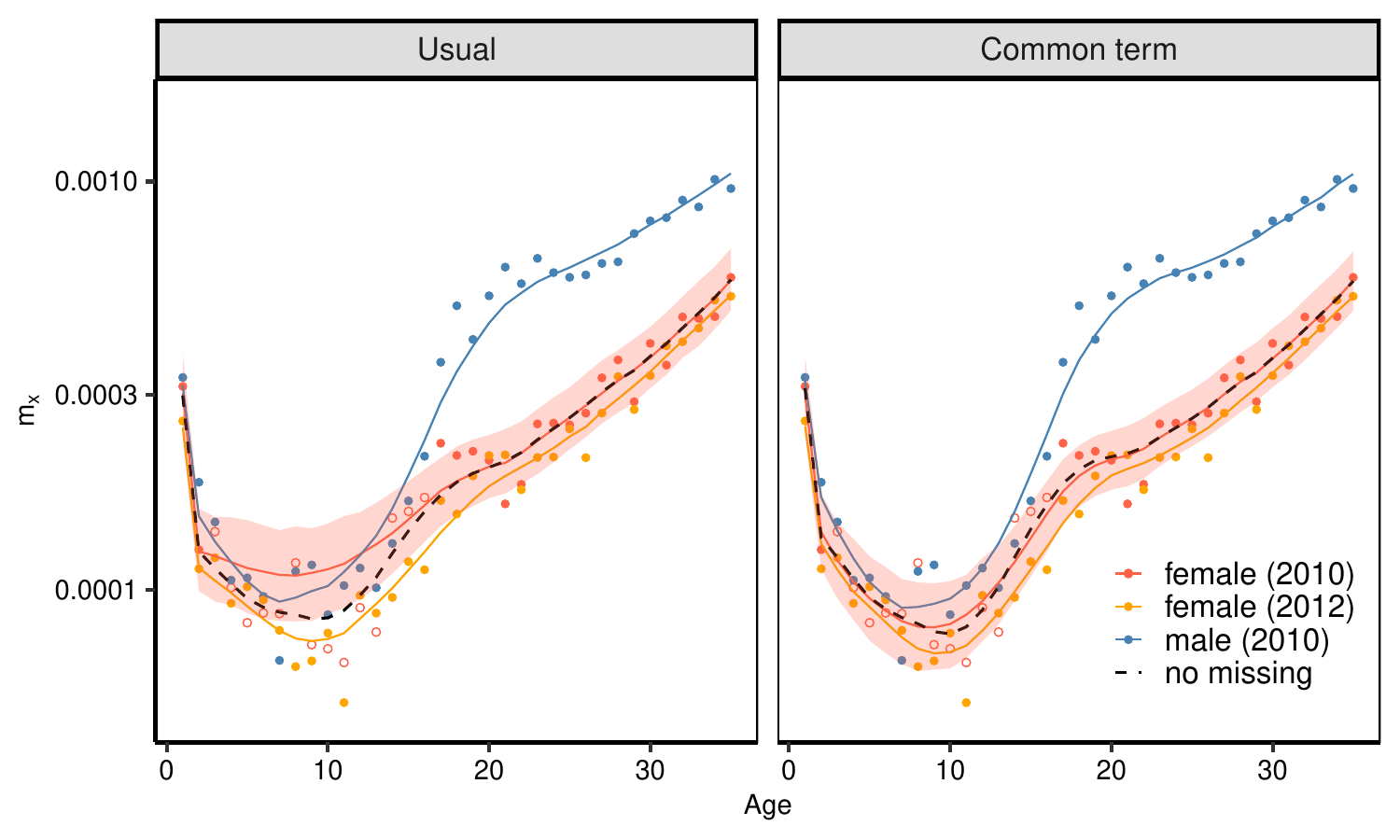} &
\includegraphics[width=0.5\textwidth]{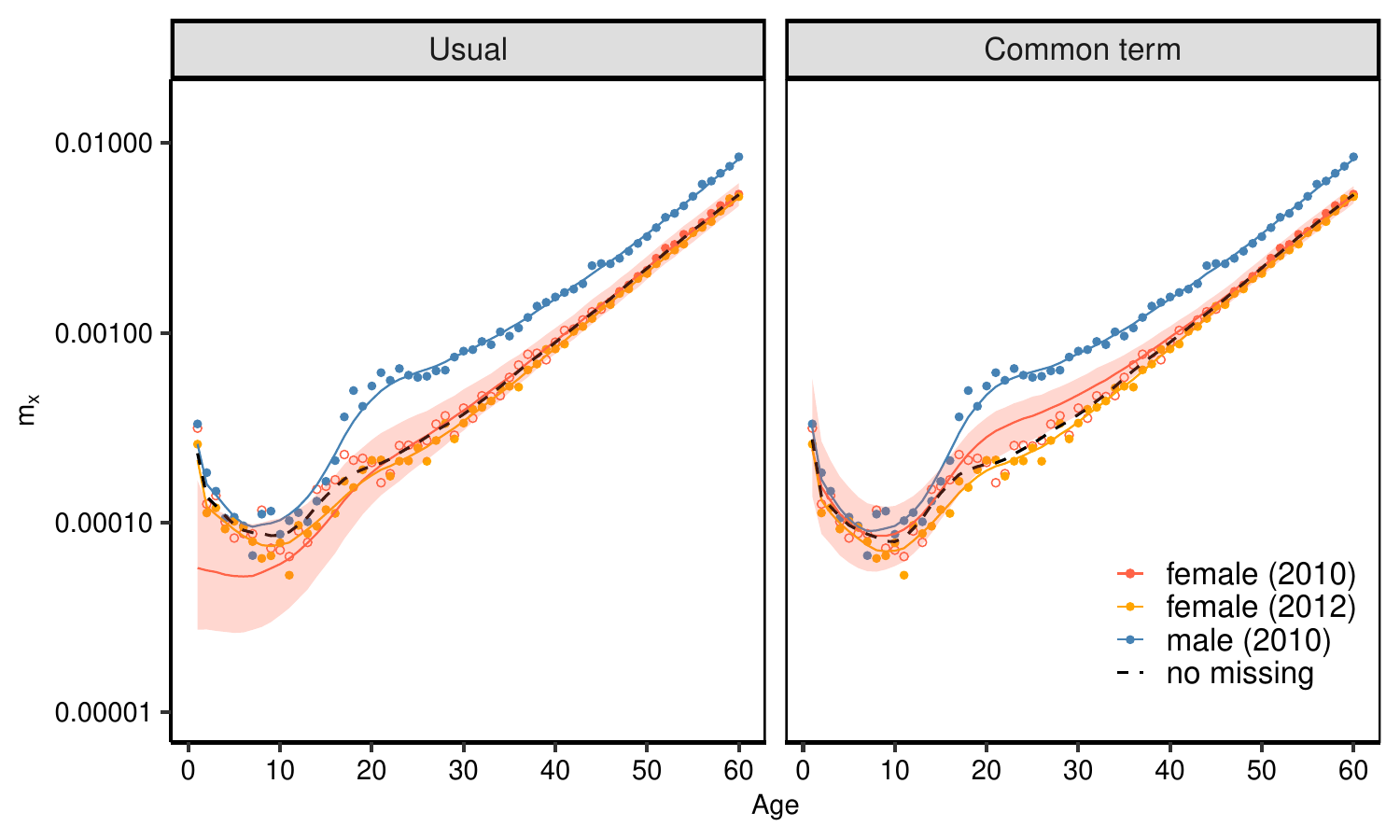}\\
scenario (c) - pooling with three populations & scenario (f) - pooling with three populations \\
\end{tabular}
\caption{Study 3 (E+W, {\bf missing data}): the DLM fits with discount factors $\delta_x$ 
 and $\delta_{1, 1:5}=0.9995$ and 95\% predictive interval - missing data for the female population (red) considering scenarios (c) and (f). Empty red dots indicate missing data. Black dashed lines represent the fit considering the complete dataset for female (2010).  }
\label{fig:fig13}
\end{figure}

\begin{figure}[H]
\centering
\begin{tabular}{c c}
\includegraphics[width=0.5\textwidth]{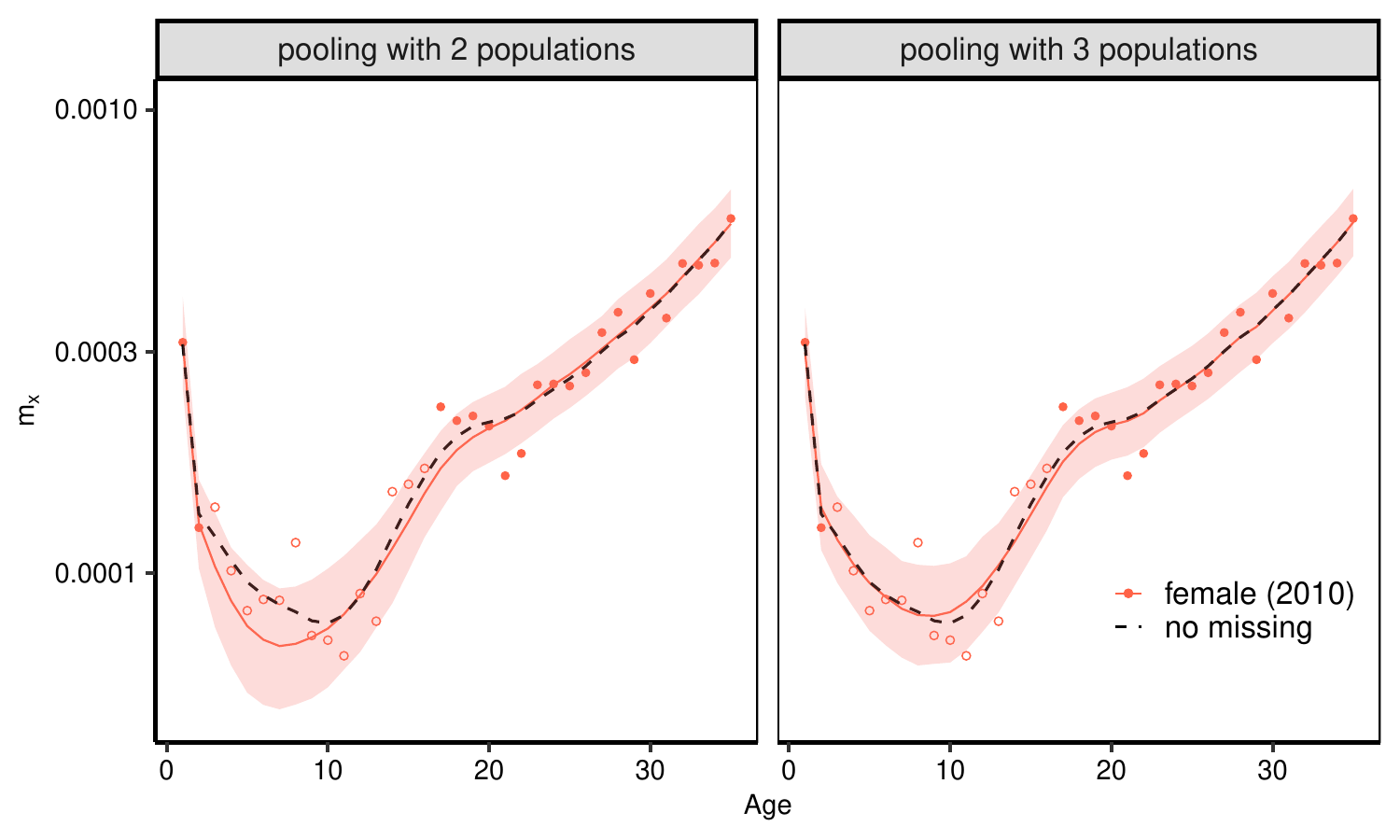} &
\includegraphics[width=0.5\textwidth]{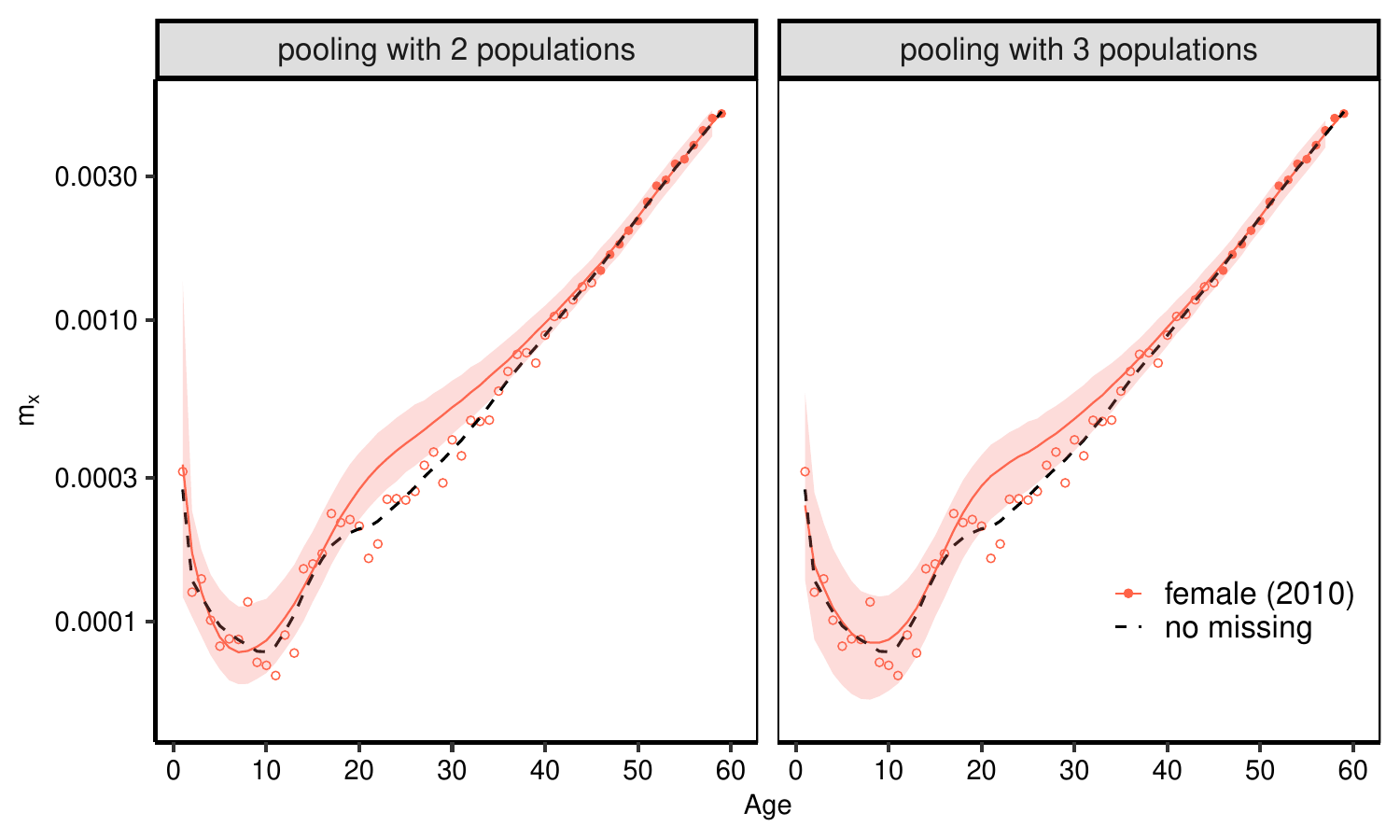}\\
scenario (c) & scenario (f)  \\
\end{tabular}
\caption{Study 3 (E+W, {\bf missing data}, {\bf joint model with common term}): comparison between pooling with two populations and three populations. the DLM fits with discount factors $\delta_x$ 
 and $\delta_{1, 1:5}=0.9995$ and 95\% predictive interval - missing data for the female population considering scenarios (c) and (f). Empty red dots represent the missing data for female (2010). Black dashed lines represent the fit considering the complete dataset.}
\label{fig:fig14}
\end{figure}

Figure \ref{fig:fig14} displays the fitted curve for females in 2010, comparing pooling with a single population (male, 2010) and pooling with two populations (male, 2010 and female, 2012) considering the joint model with common term. In scenario (c), the model that integrates data from both populations produces a fit identical to the scenario without missing data (dashed lines). When considering only male population data to fill gaps in the female population, a divergence in the curve is observed at younger ages. However, the additional information from the female (2012) data results in curves that more accurately represent the female mortality pattern. For scenario (f), the 95\% credible predictive intervals capture the observed raw mortality rates, indicating that, even with a higher percentage of missing data, the model accounts for the variability in mortality and provides reliable estimates. Incorporating additional populations that exhibit patterns similar to the population with missing data can contribute to a better fit of the scarce data subpopulation.

\newpage
\section{Conclusions}\label{sec7}

We proposed a flexible modelling approach for mortality graduation based on dynamic models, designed to induce correlation and smoothness across the age domain, while capturing subpopulation patterns even when data are scarce. By incorporating information from related subpopulations, our model effectively prevents curve crossing, resulting in more realistic mortality estimates.

Traditional univariate approaches often lead to the crossing of mortality curves, especially at older ages, which is unrealistic, particularly when comparing male and female populations. Such methods can also produce high uncertainty in extrapolated ages. To address these issues, we applied joint modelling of mortality rates, which inherently links the trajectories of mortality probabilities for both sexes. This approach avoids the unrealistic divergence or cross-over that may occur when models are treated independently, providing more coherent and realistic estimates, particularly at older ages where convergence is expected.

In cases of limited data availability, the proposed common term model effectively captures joint patterns and improves extrapolations, outperforming conventional joint models that may yield poor predictions. A key contribution of this work is the introduction of an age-varying factor, which adjusts smoothness based on the credibility of the data used for model fitting. This enhancement significantly improves model performance, especially in subnational mortality estimation, where variability in mortality patterns is usual. By allowing smoothness to vary according to data credibility, the model achieves more accurate and reliable estimates of local mortality behaviours.

Future research could focus on optimizing the selection of auxiliary series to improve subpopulation mortality estimates when data is scarce. Developing hierarchical models that integrate multiple auxiliary sources and adaptive methods that adjust weights according to data quality could enhance robustness. Additionally, performing sensitivity analysis to evaluate the impact of different auxiliary choices would provide valuable insights for reliable mortality graduation. Furthermore, extending our modelling framework based on DLM smoothers to incorporate regional or demographic information is a natural progression, taking advantage of the hierarchical Bayesian formulation of our proposal to address heterogeneity.

\newpage
\appendix

\renewcommand{\thesection}{\Alph{section}}
\renewcommand{\thetable}{\Alph{section}.\arabic{table}}
\renewcommand{\thefigure}{\Alph{section}.\arabic{figure}}
\setcounter{equation}{0}
\renewcommand{\theequation}{\Alph{section}.\arabic{equation}}

\setcounter{section}{0}
\setcounter{figure}{0}
\setcounter{table}{0}

\section{Posterior distribution and inference procedure}\label{apA}

We follow the Bayesian approach for inference, predictions, and model comparisons based on the joint posterior distribution.  In this appendix, we outline the prior specification and the steps for obtaining posterior samples. 


Let  $\boldsymbol{\theta}_{x_1: \vartheta} $ and  $ {\bf V} $ be the parameters of the proposed model presented in Section \ref{sec3}, where $ \boldsymbol{\theta}_{x_1: \vartheta} = (\boldsymbol{\theta}_{x_1}, \ldots, \boldsymbol{\theta}_{\vartheta})'$. For the covariance matrix $ {\bf V} $, we assign a vague prior distribution for $ \boldsymbol{\Phi} = {\bf V}^{-1}$ which follows a Wishart distribution with $ {\bf S}_{0_{J \times J}} $, a symmetric positive-definite correlation matrix and scale parameter $v_0 $ (with $ v_0 > (\vartheta - 1)/2 $), i.e.,
$\boldsymbol{\Phi} \sim \mathcal{W}(v_0, {\bf S}_0)$.
According to \cite{petris2009dynamic} (Chap. 4), it is convenient to express the hyperparameters of the Wishart distribution as \( v_0 = (d_0 + 1)/2 \) and \( {\bf S}_0 = \frac{1}{2}V_0 \), where \( V_0 = (d_0 - 2) \times s_0 \). The parameter \( d_0 \) controls the weight that \( {\bf S}_0 \) has in the updating process and can be weak when \( d_0 \) is close to 2, or strong when \( d_0 \) is large. For the presented studies, we set a vague prior by considering \( d_0 = 3 \) and \( s_0 = 0.01 \times \text{diag}(J) \).

Following Bayes’ theorem, the posterior distribution of $\left(\boldsymbol{\theta}_{x_1: \vartheta},{\boldsymbol{\Phi}} \right)$,
given the log mortality rate, ${\bf y}_x= \left({ y}_x^{(1)}, \ldots,{ y}_x^{(J)}\right)'$, 
$j= 1, \ldots, J$ and $x= x_1, \ldots, \vartheta$, is proportional to
{\begin{eqnarray}\label{eq:post} \nonumber
   p\left( \boldsymbol{\theta}_{x_1:
     \vartheta}, \boldsymbol{\Phi} \mid \mathbf{y}\right) &\propto&  \prod_{x=x_1}^{\vartheta}  {\mathcal{N}_{J}}\left (\mathbf{y}_x; {\bf F}_x\boldsymbol{\theta}_x, \boldsymbol{\Phi}^{-1} \right)  
    {\mathcal{N}_{p}}(\boldsymbol{\theta}_x ; {\bf G}_x \boldsymbol{\theta}_{x-1},{\bf W}_x)  \hspace{0.1cm}{\mathcal{N}_{p}}(\boldsymbol{\theta}_{x_0}; {\bf m}_{x_0},{\bf C}_{x_0}) \\ 
    & \times & {\mathcal{W}} (\boldsymbol{\Phi}; v_0, {\bf S}_0), 
\end{eqnarray}
\noindent where ${\mathcal{N}_{p}}(\cdot; A,B)$ denotes the density function of a $p$-variate multivariate Normal distribution with mean A and covariance matrix B and ${\mathcal{W}(\cdot; c, {D})}$ denotes a Wishart with parameters $c$ and matrix $D$. See that ${\mathcal{N}_{p}}(\boldsymbol{\theta}_{x_0}; {\bf m}_{x_0},{\bf C}_{x_0})$ is  the density for the initial subjective information at age $x_0$.
For ${\bf W}_x$ it is possible to use discount factor subjectively evaluates, controlling the loss of information as shown in Section \ref{sec3.2}. For more details see \cite{West97}.

Notice that the resulting posterior distribution does not have a closed form, so we employ Markov Chain Monte Carlo (MCMC) methods \citep{gamerman} to generate samples from this distribution. Specifically, we obtain posterior samples thought a Gibbs sampler scheme.



Below a general description of the MCMC algorithm to sample from the posterior full conditional distribution of the mortality dynamical linear model. Let ${\bf D}_x = \left\{ {\bf y}_x, {\bf D}_{x-1}\right\}$ the total information obtained up to age $x$. Then,\\

\noindent {\it Step 1:} Sample $(\boldsymbol{\theta}_{x_0:\vartheta}^{(k)} \mid \boldsymbol{\Phi}^{(k-1)}, {\bf D}_x)$, using forward filtering backward sampling (FFBS);\\

\noindent {\it Step 2:} Sample $(\boldsymbol{\Phi})$ from their posterior full conditional distribution using Gibbs Sampling (equation \ref{eq:eq1apA}). \\

See that the marginal posterior of $\boldsymbol{\Phi}$ is given by:
\begin{eqnarray}\nonumber \label{eq:eq1apA}
    p(\boldsymbol{\Phi} \mid {\bf y}, \cdot) & \propto& f({\bf y} \mid \boldsymbol{\Phi}, \cdot )\pi(\boldsymbol{\Phi})\\ \nonumber
     &\propto & (\det \boldsymbol{\Phi})^{\vartheta/2} \exp \left\{- \frac{1}{2}  \text{tr} \left( SS_{y}\boldsymbol{\Phi} \right)\right\} (\det \boldsymbol{\Phi})^{v_0 - (J+1)/2} \exp\left\{ - \text{tr}({\bf S}_0 \boldsymbol{\Phi})\right\}\\ 
      &\propto& (\det \boldsymbol{\Phi})^{\vartheta/2+v_0 - (J+1)/2}\exp\left\{ -
      \text{tr} \left(\left(\frac{1}{2} SS_{y} +{\bf S}_0 \right) \boldsymbol{\Phi} \right)\right\},  
\end{eqnarray}
\noindent where $\text{tr}$ represents the trace of the matrix and $SS_y = \sum_{x=x_1}^{\vartheta} ({\bf y}_x - {\bf F}_x \boldsymbol{\theta}_x)({\bf y}_x - {\bf F}_x \boldsymbol{\theta}_x)'$. Then, 

$$\boldsymbol{\Phi} \mid {\bf y}, \cdot \sim \mathcal{W} \left( v_0 + \frac{\vartheta}{2}, {\bf S}_0 + \frac{1}{2} SS_y \right). $$



\subsection{Forward Filtering Backward Sampling}

 Following \cite{Sylvia1994} and \cite{Carter1994}: \\
 
{\textit{Forward filtering equations:} 
\begin{itemize}
\item[a.] Posterior distribution at age $x_0$: $
\boldsymbol{\theta}_{x_0}| {\bf D}_{x_0} \sim \mathcal{N}({\bf m}_{x_0},{\bf C}_{x_0}); 
$

\item[b.] Prior distribution at age $x$: $\boldsymbol{\theta}_{x} | {\bf D}_{x-1} \sim \mathcal{N}({\bf a}_{x},{\bf R}_{x}),$
\noindent
with ${\bf a}_{t} = {\bf G}_{x}\textbf{m}_{x-1}$ and ${\bf R}_{x}={\bf G}_{x}{\bf C}_{x-1}{\bf G}'_{x}+{\bf W}_{x} ;$

\item[c.] One step ahead prediction:
$ {\bf Y}_{x}| {\bf D}_{x-1} \sim \mathcal{N}({\bf f}_{x},{\bf Q}_{x}), $
\noindent
with ${\bf f}_{x} = {\bf F}'_{x}{\bf a}_{x}$ and ${\bf Q}_{t}={\bf F}'_{x}{\bf R}_{x}{\bf F}_{t}+{\bf V}$, 

\item[d.] Posterior distribution at age $x$: $\boldsymbol{\theta}_{x}|\textbf{D}_{x} \sim \mathcal{N}({\bf m}_{x},{\bf C}_{x}),$
\noindent
with ${\bf m}_{x} = {\bf a}_{x}+ {\bf A}_{x}{\bf e}_{x}$ and ${\bf C}_{x}={\bf R}_{x}- {\bf A}_{x}{\bf Q}_{x}{\bf A}'_{x}$ and ${\bf A}_{t}={\bf R}_{t}{\bf F}_{x}{\bf Q}_{x}^{-1}$, ${\bf e}_{x}={\bf y}_{x}-{\bf f}_{x}$ the forecast error.
\end{itemize}

\textit{Backward Sampling equations:}\\

 This step is computed retrospectively, using the following decomposition:
 $$p(\boldsymbol{\theta}_{x_0},...,\boldsymbol{\theta}_\vartheta|{\bf D}_\vartheta)=p(\boldsymbol{\theta}_{\vartheta} | {\bf D}_\vartheta)\prod_{x=x_0}^{\vartheta}p(\boldsymbol{\theta}_x|\boldsymbol{\theta}_{x+1},{\bf D}_x).$$

 From Bayes Theorem, for $x=\vartheta-1,...,x_0$:
 $$ p(\boldsymbol{\theta}_x|\boldsymbol{\theta}_{x+1},\textbf{D}_\vartheta) \propto p(\boldsymbol{\theta}_{x+1}|\boldsymbol{\theta}_{x},\textbf{D}_\vartheta)p(\boldsymbol{\theta}_x|\textbf{D}_\vartheta),  $$
 with $\boldsymbol{\theta}_x|\boldsymbol{\theta}_{x+1},\textbf{D}_\vartheta \sim \mathcal{N}({\bf h}_x,{\bf H}_x)$. Then,
 $${\bf h}_x = {\bf m}_x+{\bf C}_x {\bf G}'_{x+1}{\bf R}_{x+1}^{-1}(\boldsymbol{\theta}_{x+1}-{\bf a}_{x+1}),$$ 
 $${\bf H}_x = {\bf C}_x - {\bf C}_x {\bf G}'_{x+1}{\bf R}_{x+1}^{-1}{\bf G}_{x+1}{\bf C}_x,$$ with ${\bf h}_{\vartheta} = {\bf m}_{\vartheta}$, ${\bf H}_{\vartheta} = {\bf C}_{\vartheta}$ being the initial values. \\

 For ${\bf W}_x$ we consider a discount factor $\delta_x \in (0,1)$ subjectively evaluated, controlling the loss of information as seen in Section \ref{sec3.2}. In this case ${\bf R}_x$ is rewritten according to a discount factor $\delta_x$ such as ${\bf W}_x = \frac{1- \delta_x}{\delta_x} {\bf G}_x {\bf C}_{x-1} {\bf G}'_x$.

\section{Model Comparison Criteria}\label{apB}
We regard the  mean square error, mean absolute error and width of credibility interval as measures for model comparison.

\subsection{Mean Square Prediction Error (MSPE)}
The mean squared prediction error measures the expected squared distance between the predictive value and the true observed value (validation observation).
\begin{equation}\label{eq:mspe}
\text{MSPE}(y, y^{pred}) = \frac{1}{\vartheta-x_0}\sum_{x=x_1}^{\vartheta} \left(y_{x} - y_{x}^{pred}\right)^2,
\end{equation}
\noindent where $z$ denotes  the  validation data, $z^{pred}$ the predictive value and $\vartheta$ is the maximum age in the validation data.

\subsection{Mean Absolute Prediction Error (MAPE)}
The mean absolute prediction error (MAPE) measures the average absolute error of predictions between the predictive value and the true observed value (validation observation).

\begin{equation}\label{eq:mape}
\text{MAPE}(y, y^{pred}) = \frac{1}{\vartheta-x_0} \sum_{x=x_1}^{\vartheta} \left|y_x - y_{x}^{pred}\right|,
\end{equation}

\noindent where lower MAPE values indicate more accurate predictions, while higher values suggest less accurate model performance.

\subsection{Width of Credibility Interval}
The Width of Credibility Interval (WCI) is defined based on credible intervals and is given by
\begin{equation}\label{eq:WCI}
WCI(u,l) = (u-l),
\end{equation}
\noindent where $l$ and $u$ represent for the forecaster quoted $\frac{\psi}{2}$ and $1-\frac{\psi}{2}$ quantiles based on the predictive distribution and $z$ is the validation observation. If $\psi=0.05$ the resulting interval has 95\% credibility.

\bibliographystyle{apalike}
\bibliography{sample}

\end{document}